\documentclass[a4paper]{article}
\usepackage[top=80pt,bottom=85pt,left=40pt,right=40pt]{geometry}
\usepackage{amssymb,amsmath,xypic}
\usepackage{bbm}
\usepackage{slashed}
\usepackage{graphicx,color,xcolor,subfigure}
\usepackage{float}
\usepackage{sectsty}
\usepackage{comment}
\usepackage{tikz}
\usepackage{braket}
\usepackage{soul}
\usepackage{cite}
\usepackage[debug,pageanchor=false]{hyperref}
\definecolor{link}{rgb}{.8,.15,.1}
\hypersetup{colorlinks=true,linkcolor=link,citecolor=link,urlcolor=link,linktocpage}

\makeatletter
\@addtoreset{equation}{section}
\makeatother

\newcommand{\beq}{\begin{equation}}
\newcommand{\eeq}{\end{equation}}
\newcommand{\bea}{\begin{eqnarray}}
\newcommand{\eea}{\end{eqnarray}}
\newcommand{\nn}{\nonumber}

\def\ds{\displaystyle}
\def\bea{\begin{array}{c}}
\def\ea{\end{array}}
\def\be{\begin{equation}\bea\ds}
\def\ee{\ea\end{equation}}
\def\bee{\begin{equation}\begin{array}{rcl}\ds}
\def\eee{\end{array}\end{equation}}

\def\nn{\nonumber}

\def\nn{\nonumber}

\newcommand\Trule{\rule{0pt}{2.5ex}}
\newcommand\Brule{\rule[-1.ex]{0pt}{0pt}}

\begin{document}
\begin{titlepage}

\begin{center}

\vskip .5in 
\noindent

{\Large \bf{On generalised D1-D5 near horizons and their spectra}}

\bigskip\medskip

Mariana Lima$^{a,b,c}$\footnote{m.limalp@gmail.com},  Niall T. Macpherson$^{d,e}$\footnote{ntmacpher@gmail.com}, 	Dmitry Melnikov$^{a}$\footnote{dmitry@iip.ufrn.br}, Luis Ypanaqu\'e$^{a,f}$\footnote{luis.12yr@gmail.com} \\

\bigskip\medskip
{\small 

   $a$: International Institute of Physics, Federal University of Rio Grande do Norte,\\ 
       Campus Universit\'ario - Lagoa Nova, Natal, RN 59078-970, Brazil\\
			
\vskip 3mm
	$b$: Department of Theoretical and Experimental Physics, Federal University of Rio Grande do Norte,\\ 
	Campus Universit\'ario - Lagoa Nova, Natal, RN 59078-970, Brazil\\

\vskip 3mm
	$c$: Department of Physics, Swansea University,\\ 
	Swansea SA2 8PP, United Kingdom\\

\vskip 3mm
	$d$: Department of Physics, University of Oviedo,\\ 
	Avda. Federico Garcia Lorca s/n, 33007 Oviedo
\\
	and
\\
	$e$: Instituto Universitario de Ciencias y Tecnolog\'ias Espaciales de Asturias (ICTEA),\\ 
	Calle de la Independencia 13, 33004 Oviedo, Spain  \\


\vskip 3mm
	$f$: School of Mathematical Sciences, University of Southampton,\\
       Highfield, Southampton, SO17 1BJ, United Kingdom  \\
}
\bigskip\medskip

\vskip 1cm 

     	{\bf Abstract }
     	\end{center}
     	\noindent
	
There has been considerable recent interest in type II string compactifications on AdS$_3\times$S$^3\times$M$_{4}$ due to both new progress in the AdS$_3$/CFT$_2$ correspondence and an independent search for new ${\cal N}=(4,0)$ supergravity solutions. In this work we report a new family of small ${\cal N}=(4,0)$ preserving warped AdS$_3\times$S$^3\times $CY$_2$ compactifications of type IIB supergravity. The S$^3$ in this family is non trivially fibered over the CY$_2$. We show how these new solutions provide a IIB embedding of minimal ungauged $d=5$ supergravity coupled to an Abelian vector multiplet.  We also make a few initial steps towards the identification of dual CFT$_2$ candidates. Specifically, we study the Kaluza-Klein spectrum of the spin two and certain vector fluctuations that are expected to be dual to the stress-energy tensor, SU(2) R current and several other BPS operator families in the dual CFT$_2$.

\noindent

\vfill
\eject

\end{titlepage}

\tableofcontents


\section{Introduction}

AdS$_3$ compactifications of strings and supergravity are of great interest for the AdS/CFT correspondence because they are supposed to describe duals of 2d $1+1$-dimensional conformal field theories~\cite{Maldacena:1997re}. Strings in AdS$_3$ with NS-NS flux were initially studied in~\cite{Elitzur:1998mm,Maldacena:2000hw,Maldacena:2000kv,Maldacena:2001km}, where a lot of information was obtained, including spectra and correlation functions, but it also became clear that the dual CFTs belong to a special class. In particular, the spectrum of such theories must in general contain a continuum of states dual to long strings. It has been suspected for long time that strings in AdS$_3\times$S$^3\times{\cal M}_4$ are dual to CFTs on symmetric product orbifolds $({\cal M}_4)^N/S_{N}$, but which precise AdS$_3$ backgrounds and which precise CFTs are related by the duality has been unclear.

Part of the tests of this AdS$_3$/CFT$_2$ correspondence where performed in the strong coupling limit of the putative CFTs, which corresponds to the classical gravity limit of strings. The gravity limit allows one to deduce information about the BPS spectrum, which can be tested in the dual CFTs beyond strong coupling, because BPS states are protected against quantum corrections. For compactifications on AdS$_3\times$S$^3$ preserving ${\cal N}=4$ supersymmetry, either small (${\cal{N}}=(4,0)$) or large (${\cal{N}}=(4,4)$), the BPS spectrum was studied in the greatest generality in~\cite{deBoer:1998kjm,deBoer:1999gea}. While in general the spectra showed consistency with the world sheet analysis and with symmetric orbifold CFTs, in particular in the sector with only the Neveu-Schwarz flux turned on~\cite{Maldacena:1998bw,Kutasov:1998zh,Argurio:2000tb}, there were a few puzzles. One of them was related to large multiplicity of the BPS spectrum in geometries preserving the large superconformal algebra~\cite{Gukov:2004ym,deBoer:1999gea}. This algebra contains a SU(2)$\times$SU(2) subalgebra whose representations correspond to angular momenta $\ell_1$ and $\ell_2$ on the S$^3$ of AdS$_3\times$S$^3$ and on the S$^3$ within ${\cal M}_4$. The BPS bound in this algebra is saturated by $\ell_1=\ell_2$, but this had not been seen in the analysis \cite{deBoer:1999gea}, which relied on group theory arguments (rather than a direct computation of the spectrum in supergravity) seemingly permitting states with independent $\ell_1$ and $\ell_2$.

More recently significant progress addressing these problems has been achieved. A detailed analysis of the type IIB supergravity spectrum on AdS$_3\times$S$^3\times$S$^3\times$S$^1$~\cite{Eberhardt:2017pty}, beyond the representation theory arguments of~\cite{deBoer:1998kjm} and~\cite{deBoer:1999gea}, has demonstrated that the BPS bound $\ell_1=\ell_2$ is in fact satisfied at the supergravity level and no infinite degeneracy of the BPS states occur, and no far-fetched quantum correction is necessary to remove redundant states. Moreover it proved possible to demonstrate the precise correspondence between the tensionless limit of strings ($k=1$ unit of NS flux) on this background~\cite{Eberhardt:2019niq}, and on AdS$_3\times$S$^3\times$T$^4$~\cite{Eberhardt:2018ouy}, and CFTs on the respective symmetric orbifolds.

An obvious follow up direction is to search for more examples of exact AdS$_3$/CFT$_2$ correspondence. In this work we turn to \emph{warped} AdS$_3\times$S$^3\times$CY$_2$ compactifications of type IIB string theory that preserve the small ${\cal N}=(4,0)$ superconformal algebra. The canonical example of a solution realising this algebra is the D1-D5 near horizon \cite{Maldacena:1997re}, which preserves $(4,4)$ supersymmetry, so two copies of the small algebra of opposing chiralities. The simplest generalisation of this is the D1-D5+KK near horizon \cite{Kutasov:1998zh,Sugawara:1999qp,Larsen:1999dh,Okuyama:2005gq} which replaces $\text{S}^3\!\to\text{S}^3/\mathbb{Z}_k$ in the above breaking supersymmetry to ${\cal N}=(4,0)$. 

In recent years there has been some renewed effort to construct small ${\cal N}=(4,0)$ solutions in type II supergravities \cite{OColgain:2010wlk,Lozano:2015bra,Couzens:2017way,Lozano:2019emq,Lozano:2019jza,Lozano:2019zvg,Lozano:2019ywa,Faedo:2020lyw,Dibitetto:2020bsh,Faedo:2020nol,Zacarias:2021pfz,Couzens:2021veb,Couzens:2020aat,Macpherson:2022sbs,Lozano:2022ouq}, of particular interest to us here is the SU(2)-structure classification of \cite{Lozano:2019emq}: The main focus of this work was to construct local solutions in massive IIA on AdS$_3\times$S$^2\times$CY$_2$, that were later used to construct holographic duals to ${\cal N}=(4,0)$ linear quivers \cite{Lozano:2019jza,Lozano:2019zvg,Lozano:2019ywa}. However \cite{Lozano:2019emq} also found a pronounced generalization of the D1-D5 near horizon: This is a warped AdS$_3\times $M$_7$ class with $\text{S}^3\!\hookrightarrow\text{M}_7\!\rightarrow\text{CY}_2$ and solutions in one-to-one correspondence with the solutions of a generalised Laplace equation on CY$_2$ -- it preserves small ${\cal N}=(4,0)$. The presence of a non trivial CY$_2$ dependent warp factor means that one is not constrained to only considering $\mathbb{T}^4$ and K3, one can also achieve a bounded M$_7$ with non compact CY$_2$s bounded to some finite sub-region by D5 branes and/or O5 planes. The first example of such a solution was found in \cite{Macpherson:2018mif} namely a solution on AdS$_3\times $S$^3\times \mathbb{R}^4$ bounded between an O5 and D5s but with no fibration turned on. Part of the motivation of this work is to construct more general solutions of this type, with non trivial fibration\footnote{note that \cite{Couzens:2021veb} has some overlap with this goal, albeit for related IIA solutions without such a fibration.}, another is to make some headway into understanding the CFT duals of such solutions.

In this work we present new solutions with S$^3$ non trivially fibered over CY$_2$. We consider the case where CY$_2= \mathbb{R}^4$ in some detail, constructing solutions bounded between D5 brane and O5 plane singularities,  O5-O5 singularities and between an O5 and a regular zero. We also consider the case where CY$_2= \mathbb{T}^4$, and derive  PDEs  defining solutions for which CY$_2$ contains a U(1) isometry but is otherwise completely generic. In addition to constructing new AdS$_3$ vacua, we show that the class in which they reside can be generalised to provide a new type IIB embedding for all solutions of minimal  ungauged supergravity in 5 dimensions  coupled to an additional Abelian vector multiplet.  

With a set of new supersymmetric backgrounds in hand, the first step in identifying the dual CFT is identifying the spectrum of strings in such backgrounds. While this turned out to be possible for the unwarped solutions dual to symmetric orbifold CFTs of~\cite{Eberhardt:2017pty,Eberhardt:2018ouy,Eberhardt:2019niq}, this is a hard problem in general. A more modest goal is to determine the spectrum of light operators using the supergravity approximation of string theory, in which case the problem is reduced to finding the Kaluza-Klein spectrum of the respective AdS$_3$ compactification. Even this simpler problem is generally out of reach, so only a handful of examples are known in their full glory, e.g. the result of~\cite{Eberhardt:2017fsi} for AdS$_3$ and~\cite{Biran:1983iy,Kim:1985ez,Ceresole:1999zs} in higher dimensions. New hopes are connected with the method of Kaluza-Klein spectrometry~\cite{Malek:2019eaz,Malek:2020yue}, based on the formalism of  exceptional field theory. The new methods seem to be rather powerful, extending beyond supersymmetric backgrounds. Some  AdS$_3$ spectra for simple solutions have been recently revisited and corrected in~\cite{Eloy:2020uix}. However it is our understanding that such spectrometry methods are only fully developed for AdS$_3$ solutions of $d=6$ gauged supergravity at this time, with a full mapping of $d=10$ fields to ``exceptional geometry'' still lacking for AdS$_3$. Even if this were not the case the situation here is complicated by the presence of D brane and O plane sources whose fluctuations need to be taken into account -- this was not yet attempted by the exceptional field theory approach. 

Our take on the Kaluza-Klein spectrum of a new family of $(4,0)$ solutions in this work is by analysing its most protected species, included in the spin two multiplet\footnote{Unless explicitly specified the spin of the mode is its spin with respect to the isometry group of AdS$_3$ defined as $s=|h-\bar h|$, where $h$ and $\bar h$ are the left-moving and right-moving conformal weights.}, that is the spectrum of the stress-energy tensor, R current and associated BPS families. The graviton, or the spin two fluctuation of the metric, is a particularly simple mode as it happens to decouple from fluctuations of other fields and sources. This fact was proven for backgrounds that are warped products containing a maximally symmetric 4-manifolds in~\cite{Bachas:2011xa}. A particular generalisation of that analysis applies here as well. We also show that spin one fluctuation dual to the triplet of R currents of $(4,0)$ SUSY is a relatively simple mode, although not directly obtained from a fluctuation of a single type IIB field. Both observations are consequences of SUSY that places R current and the stress-energy modes in a single short multiplet. Consequently, the spin one equation can as well be generalized to a generic set of backgrounds, as is the case of the spin two mode.

In this work we only discuss the R current mode for the simplest representative of the class of new supergravity solutions, leaving generalizations for another paper~\cite{inprep}. This mode comes from a coupled fluctuations of the metric and the Ramond-Ramond 2-form potential that are bi-vectors with respect to AdS$_3$ and S$^3$. The fact that the R current mode requires perturbations of two supergravity fields, allows us to detect a few additional decoupled spin one modes. Together with the spin two perturbation of the metric along AdS$_3$ directions the spectrum of the above perturbations reveals the expected BPS branches, containing the stress-energy tensor with conformal dimension $\Delta=h+\bar h=2$, the R current $\Delta=1$; the corresponding towers of heavier BPS operators labeled by the S$^3$ spin $\ell_1$, $(\Delta=\ell_1+2,\,s=|h-\bar h|=2)$ and $(\Delta=\ell_1,\,s=1)$, which are the highest and the lowest members of the short spin two multiplets; and three additional BPS families of vectors, two with $(\Delta=\ell_1+2,\,s=1)$ and one with $(\Delta=\ell_1+4,\,s=1)$.

The analysis of the spectrum requires an appropriate choice of boundary conditions for the fluctuations along CY$_2$, and in particular in the compact direction transverse to S$^3$. In this work we applied the general set of conditions previously used in the literature, e.g.~\cite{Passias:2016fkm,Passias:2018swc}. Such conditions are justified by the variational principle, which is compatible with either fixing the fluctuations at the boundaries (Dirichlet case) or making them satisfy a canonically conjugate boundary condition (Neumann case). In fact, the conditions used in~\cite{Passias:2016fkm,Passias:2018swc} assume the strong form of Dirichlet boundary conditions (fluctuations vanish), for regular boundary points. The strong form of boundary conditions implies that the eigenvalues of the associated Sturm-Liouville problem are non-negative. We find then that in the considered set of backgrounds only the modes satisfying the Neumann boundary conditions can saturate the BPS bound of the superconformal algebra. We also demonstrate, how the multiplicity of the BPS spectrum, mentioned in the case of AdS$_3\times$S$^3\times$S$^3\times$S$^1$, is avoided in the supergravity analysis with the above boundary conditions: BPS states of the superconformal algebra with general values of $\ell_2$ (spin on S$^3$ within CY$_2$ or S$^3\times$S$^1$) violate the non-negativity of the eigenvalues.

We also checked the spectrum of non-BPS states and found two qualitatively different pictures in the O5-O5 and O5-D5 scenarios. In the former case one obtains a bona fide discrete spectrum, asymptotically linear for large values of the respective quantum number. In the latter case, there is only a finite number of eigenstates, that number being a function of the moduli of the background, such as the number $N_5$ of the original D5 branes. The discrete spectrum is bounded from above by a continuum cut, at which the fluctuations become unnormalisable. The cut moves to infinity as one interpolates between the O5-D5 and O5-O5 situations.

There are a few important points that one should keep in mind in the discussion of the spectrum. First, D-branes and O-planes act as sources for the bulk supergravity fields. Consequently, when the background fields are perturbed, fluctuation of the sources must be included, or their absence must be justified. In the case of the spin two mode decoupling from the sources was checked in~\cite{Bachas:2011xa}. The backgrounds of the present work fit in the general class discussed by that paper. Spin one fluctuations considered here also decouple from the sources. This will be shown in a separate work for a more general class of backgrounds~\cite{inprep}. Second, solutions with D-brane and O-plane sources are often singular, so the supergravity description should break down in the vicinity of such objects, and it is not known in general, how this problem can be addressed. Here we assume that for the fact of the backgrounds being supersymmetric the supergravity results will still be valid after appropriate singularity resolving corrections are taken into account. This must be true as long as the BPS states are considered, and an interesting question is whether the observed pecularities of the non-BPS states in the O5-D5 systems would survive in the full picture as well.

This paper is organised as follows. In section~\ref{sec:backgrounds} we introduce a family of ${\cal N}=(4,0)$ AdS$_3\times$S$^3\times $CY$_2$ backgrounds generalising  previous results, which are reviewed in section~\ref{sec:basicsol}. More general new solutions are reviewed in section~\ref{sec:fibredsols}, where we consider the cases of compactifications on global $\mathbb{R}^4$ (section~\ref{sec:globalR4}), four-torus (section~\ref{sec:4torus}) and on Calabi-Yau manifolds with a circle fibration (section~\ref{sec:circlefibration}). All the above backgrounds are supported by a RR 3-form flux. In section~\ref{sec:Sdual} we provide a new emedding of  minimal $d=5$ supergravity, coupled to an Abeilian vector multiplet, into type IIB supergravity. In the following sections we present partial results on the spectrum of linear perturbations over the discussed family of backgrounds. In section~\ref{sec:spin2} we study the spin two  perturbations of the metric. The analysis of the spectrum of the spin two mode is performed for the known solutions of section~\ref{sec:basicsol} (section~\ref{sec:spin2noF}) and for the new family~(section~\ref{sec:spin2F}). The purpose of section~\ref{sec:simplespec} is to show the difference of the BPS spectrum in the case of small and large ${\cal N}=4$ superalgebras and explain how boundary conditions solve the problem of degeneracy. Section~\ref{sec:spin1} contains the analysis of the spin one mode that is expected to be dual to the SU(2) R-current and a few coupled spin one supergravity modes. Section~\ref{sec:stringsAdS3} discusses strings on orientifolds and explains the reduction of the spectrum in the case of AdS$_3\times$S$^3\times$T$^4$. We conclude in section~\ref{sec:conclusions}. The paper also contains a number of appendices. Appendix~\ref{sec:gency2withu1appendix} explains the existence of two classes of two-dimensional Calabi-Yau manifolds with a U(1) isometry. Appendices~\ref{sec:smallN4} and~\ref{sec:largeN4} review the representation theory of small and large ${\cal N}=4$ superalgebras respectively. Finally, in Appendix~\ref{sec:harmonics} we summarize some relevant facts about spherical harmonics on the 3-sphere.


\section{Small ${\cal N}=(4,0)$ geometries}
\label{sec:backgrounds}

In \cite{Lozano:2019emq} a new class of AdS$_3$ solutions realising small ${\cal N}=(4,0)$ in type IIB supergravity was found. Solutions in this class generalise the D1-D5 near horizon geometries, AdS$_3\times$S$^3\times $CY$_2$ for CY$_2=(\mathbb{T}^4,\text{K}3)$ via additional D5 branes and by non trivially fibering the 3-sphere over CY$_2$. All such solutions can be expressed in the string frame form
\begin{align}
ds^2&= \frac{L^2}{\sqrt{h_5}} \bigg[ds^2(\text{AdS}_3)+\frac{1}{4}ds^2(\text{S}^2)+ \frac{1}{4}D\psi^2\bigg]+ \lambda^2\sqrt{h_5}ds^2(\text{CY}_2)\,, \qquad e^{\Phi}= \frac{L^2}{c\sqrt{h_5}}\,,\nn\\[2mm]
F_3& = 2c\bigg[\text{vol}(\text{AdS}_3)+\frac{1}{8}D\psi\wedge \text{vol}(\text{S}^2)\bigg]-\frac{c}{4}D\psi\wedge {\cal F}+ \frac{c\lambda^2}{L^2}\star_4 dh_5\,.\label{eq:CYsolgen}
\end{align}
Here $(c,\lambda,L)$ are constants. The coordinate $\psi$ parametrises the U(1) of the Hopf fibration of  S$^3$, which is invariant under the action of SU(2)$_R$. The fibration over CY$_2$ is mediated by a connection 1-form ${\cal A}$, with support on CY$_2$, 
\be
D\psi = d\psi+ {\cal A}+\eta, \qquad d\eta= \text{vol}(\text{S}^2)\,.
\ee
The field strength of ${\cal A}$ must be anti-self dual for supersymmetry to hold,\footnote{Note $\star_4$ is the Hodge dual on the unwarped CY$_2$.}  
\be
\label{Fcurv}
{\cal F}= d{\cal A}, \qquad \star_4{\cal F}+ {\cal F}=0\,.
\ee
The remaining fluxes are trivial\footnote{One can however turn on the RR 1-form and NS 3-form via an SL(2,$\mathbb{R}$) transformation.}. The metric on the 2-sphere and AdS$_3$ has unit radius and the period of $\psi$ should be $4\pi k$, for $k$ an integer, when $k=1$ there is a round 3-sphere, when $k>1$ one has a $\mathbb{Z}_k$ orbifold of this. The warp factor $h_5$ has support on CY$_2$ and away from the loci of sources, must obey
\beq\label{eq:unsourcedPDE}
\nabla_{\text{CY}_2}^2h_5+ \frac{L^2}{8\lambda^2}{\cal F}^2 \ = \ 0\,,
\eeq
where we define ${\cal F}^2= {\cal F}_{ab}{\cal F}^{ab}$, with $a,b$ flat directions on the unwarped CY$_2$. Were it not for the contribution from the connection this would formally be the PDE of D5 branes back-reacted on CY$_2$,  indeed the ${\cal A}=0$ limit of~(\ref{eq:CYsolgen}) can actually be realised as a simple generalisation of the D1-D5-KK near horizon geometry (also preserving ${\cal N}=(4,0)$). The presence of the connection allows for non trivial generalisations, the near horizon realisation of these geometries is provided  (modulo T-duality on $\psi$) by \cite{Couzens:2021veb}.

The presence of the warp factor in \eqref{eq:CYsolgen} means that it is possible to obtain bounded ten-dimensional solution for Calabi-Yau 2-folds other than $(\mathbb{T}^4,\text{K3})$. This is because one no longer needs to demand that CY$_2$ is itself compact as long as the warp factor bounds it to some finite sub-region. The first example of such a solution, was given  in \cite{Macpherson:2018mif}  (actually predating \cite{Lozano:2019emq}) with the warp factor bounding $\mathbb{R}^4$ to a finite region between O5 and D5 branes, with ${\cal F}=0$. As this example suggests, arranging for such a bounded solution will in general necessitate some sources, so  \eqref{eq:unsourcedPDE} needs to be generalised to
\beq\label{eq:sourcedPDE}
\nabla_{\text{CY}_2}^2h_5+ \frac{L^2}{8\lambda^2}{\cal F}^2= \frac{1}{\sqrt{g_{\text{CY}_2}}}\sum_i Q_i \delta(\underline{x}-\underline{x}_i) \,,
\eeq 
where  $\underline{x}$ are coordinates on CY$_2$ and there are a number of sources at the loci $\{\underline{x}_i\}$. The Dirac delta functions are such that $\int_{\text{CY}_2}\frac{1}{\sqrt{g_{\text{CY}_2}}}\delta(\underline{x}) \text{vol}(\text{CY}_2)=1$. Each of these  sources could be either a stack of D5 branes, an O5 plane or a coincident combination of both. The charge of a single D5 or O5 plane are given by
\beq
Q^{\text{D5}}=- 2\kappa^2_{10}T_5= -(2\pi)^2\,, \qquad Q^{\text{O5}}=-Q^{\text{D5}}\,,
\eeq
in units where $g_s=\alpha'=1$.

In the next section we shall review the solution of \cite{Macpherson:2018mif}. In the sections that follow we shall present some totally new solutions with ${\cal F}\neq 0$.

\subsection{D5-O5 back-reacted on $\mathbb{R}^4$ with ${\cal F}=0$}
\label{sec:basicsol}

The solution of \cite{Macpherson:2018mif} is probably the most simple example of a solution within the class of \eqref{eq:CYsolgen} that goes beyond the original D1-D5 near horizon geometry~\cite{Maldacena:1997re}. For this case ${\cal A}=0$ and we take CY$_2$ to be simply $\mathbb{R}^4$, which we express in polar coordinates, such that we now have two 3-spheres. We define these via
\beq\label{eq:spheres}
ds^2(\text{S}^3_1)= \frac{1}{4}(ds^2(\text{S}^2)+ D\psi^2),~~~~ ds^2(\mathbb{R}^4)= dr^2+ r^2 ds^2(\text{S}^3_2).
\eeq
If we assume that the SO(4) symmetry of $\text{S}^3_2$ is preserved then \eqref{eq:unsourcedPDE} is solved by
\beq
\label{h5}
h_5 \ =\  b_1 +\frac{b_2}{r^2}\,,
\eeq
which is a D5 brane warp factor if $b_{1,2}>0$. If one instead assumes $b_1<0$, $b_2>0$ then the warp factor interpolates between D5 and O5 behaviours. One can make the bounded behaviour of this latter choice more explicit, and the solution easier to interpret by fixing 
\beq
b_2=-b_1=1\,, \qquad r=\cos \rho\,, \qquad \lambda^2=\frac{b}{c}L^2.
\eeq
The solution then becomes
\begin{align}
ds^2&= L^2\bigg[\frac{1}{\tan \rho}\bigg(ds^2(\text{AdS}_3)+ds^2(\text{S}^3_1)\bigg)+ \frac{b}{c}\tan \rho\bigg(\sin^2\!\rho\, d\rho^2+\cos^2\!\rho\, ds^2(\text{S}^3_2)\bigg)\bigg]\nn\\[2mm]
F_3&=2c \bigg[\text{vol}(\text{AdS}_3)+\text{vol}(\text{S}^3_1)\bigg]-2 b \,\text{vol}(\text{S}^3_2)\,,\qquad e^{\Phi} \ = \ \frac{L^2}{c\tan \rho}, \label{eq:solsimp}
\end{align}
where $\rho \in [0,\frac{\pi}{2}]$ with the metric tending to the behaviours of respectively an O5 plane or D5 branes (both extended in AdS$_3\times$S$^3_1$) at the lower and upper bounds.  This solution has an enhanced symmetry with respect to \eqref{eq:CYsolgen}, i.e. SU(2)$\times$U(1)$\ \to\ $SO(4)$\times$SO(4). Despite this, supersymmetry is still just ${\cal N}=(4,0)$, and one may actually perform Z$_k$ orbifoldings of both 3-spheres without breaking this further. 
 
Flux quantisation demands that we tune the constants $(b,c,L)$ such that
\beq\label{eq:R4fluxquantisation}
\frac{1}{(2\pi)^2}\int_{\text{S}^3_1}F_3= N_5\,, \qquad -\frac{1}{(2\pi)^6}\int_{\text{S}^3_1\times \text{S}^3_2\times {\cal I} }\star F_3= N_1\,, \qquad -\frac{1}{(2\pi)^2}\int_{\text{S}^3_2}F_3= n\,,
\eeq
are all integers, which  amounts to fixing
\beq\label{eq:quant1}
 \frac{b^2 L^4}{32 c\pi^2} = N_1\,, \qquad c= N_5\,, \qquad b=n\,.
\eeq
$N_1$, $N_5$ are associated to the number of colour D1 and D5 branes that are already present for the standard the D1-D5 near horizon. When they are taken to be large  the solution is weakly coupled every where but a tunable radi about the source D5s and O5 \cite{Macpherson:2018mif}. The charge $n$ is related to the sources, its interpretation is a bit more subtle and depends on the interpretation of the internal space. The sources we see in the solution are actually consistent with two scenarios that supergravity alone can not distinguish between: In both cases there is an O5 plane at $\rho=0$ but  at $\rho={\pi}/{2}$ there could be either a single D5 brane or a stack of 2 D5s coincident to another O5 plane. The latter option suggests an interesting possibility\footnote{We are indebted to Alessandro Tomasiello for discussions on this point.}:  One could interpret the warped CY$_2$ as a 4-sphere, with an orientifold acting on its embedding coordinates as  $(y_1,...,y_5) \to (-y_1,...,-y_4,y_5)$. This action has two fixed points at the poles $(0,0,0,0,\pm 1)$, which is the loci of the two sets of sources. As we view the internal space as a quotient of a compact space, the integrated form of \eqref{eq:sourcedPDE} (with ${\cal F}=0$) imposes that  $\sum_i Q_i=0$. Then as there are 2 O5 planes we need 2 D5 branes to cancel their charge and so $n=1$.

We can use the expression of \cite{Couzens:2017way} to compute the holographic central charge. This tells we have the following relation between the (string frame\footnote{Note that reference \cite{Couzens:2017way} uses the Einstein frame.}) metric, dilaton and central charge
\beq\label{eq:centralcharge}
ds^2= e^{2A}ds^2(\text{AdS}_3)+ ds^2(\text{M}_7)\,, \qquad c_{hol}= \frac{3}{2^4 \pi^6}\int_{\text{M}_7}e^{A-2\Phi}\text{vol}(\text{M}_7).
\eeq 
Reading the relevant quantities off \eqref{eq:solsimp} and using \eqref{eq:quant1} it is simple to show that this yields
\beq
c_{hol}= 6 N_1 N_5,
\eeq 
consistent with what one expects from a small ${\cal N}=(4,0)$ CFT of level $k=N_1 N_5$.

\section{Solutions with non trivial ${\cal F}$  }
\label{sec:fibredsols}

In this section we shall construct some new solutions of~(\ref{eq:CYsolgen}) type with non trivial ${\cal F}$. We begin by giving some details of this new ingredient.

A Calabi-Yau 2-fold can be defined in terms of three self dual 2-forms $(J_1,J_2,J_3)$ that obey
\beq
J_a\wedge J_b\ =\ 2 \delta_{ab}\,\text{vol}(\text{CY}_2)\,,\qquad  dJ_a \ = \ 0\,, \qquad a,b=1,2,3.
\eeq
The anti self-duality of the 2-form ${\cal F}$ introduced in equations~(\ref{eq:CYsolgen}) and~(\ref{Fcurv}) is equivalent to the conditions
\beq\label{eq:primative}
J_a\wedge {\cal F}=0\,, \qquad a=1,2,3,
\eeq
which makes it  a primitive $(1,1)$-form by definition.  In general there exists a canonical frame  on CY$_2$ with respect to which
\beq
J_1=e^{12}+e^{34}\,, \qquad J_2=e^{13}-e^{24}\,, \qquad J_2=e^{14}+e^{23}\,,
\eeq
where $e^{ij}=e^i\wedge e^j$. In such a frame we can in general expand ${\cal F}$ in terms of  three functions $f_a$ with support on CY$_2$ as
\beq
\label{Fansatz}
{\cal F}= f_1 (e^{12}- e^{34})+f_2( e^{13}+ e^{24})+f_3( e^{14}- e^{23})\,,\qquad {\cal F}^2=4 (f_a)^2\,.
\eeq
Thus to construct a solution with S$^3$ fibered over a specific CY$_2$, the first step is to find a set of functions such that $d{\cal F}=0$. This is quite easy to do when CY$_2$ is locally $\mathbb{R}^4$, so we can take 
\beq
e^{i} \ =\  dx^{i}\,, \qquad i\ =\ 1,\ldots,4\,.
\eeq
Generically this gives rise to 4 PDEs which are rather non trivial, they imply $f_a$ are harmonic, but are more restrictive than this. However ${\cal F}$ is closed when one simply takes $f_a$ constant, and this already suffices to have a non trivial fibration.  So we have a concrete proposal for a closed ${\cal F}$ that can be applied when CY$_2$ is globally either $\mathbb{R}^4$, $\mathbb{T}^4$ or some quotient thereof. In the former case it is convenient to write $x_i$ in polar coordinates, then ${\cal F}$ decomposes in terms of the radial coordinate and the SU(2) invariant forms. In these distinct cases we can take the connection 1-form to be 
\beq
{\cal A} \ = \ \left\{\begin{array}{l} \ds
\frac{r^2}{4}(c_1 L_1+c_2 L_2+c_3 L_3)\,, \qquad  \text{for}\quad \mathbb{R}^4=(r,\text{S}^3),\qquad ds^2(\text{S}^3)=\frac{1}{4}(L_a)^2,\\[3mm]
\ds c_1(x_{[1} dx_{2]}-x_{[3} dx_{4]})+ c_2 (x_{[3} dx_{1]}-x_{[2} dx_{4]})+c_3 (x_{[2} dx_{3]}-x_{[1} dx_{4]})\,,\qquad  \text{for}\quad \mathbb{T}^4,\end{array}\right.
\eeq
where $L_a$  are a set of SU(2) left invariant 1-forms obeying $dL^a=\frac{1}{2}\epsilon_{abc}L^b\wedge L^c$ and $c_{1,2,3}$ are constants (they are $f_a$ up to signs).  In either case what remains to be solved is the generialised Laplace equation
\beq\label{eq:PDElocalR4}
\nabla^2_{\text{CY}_2}h_5+  {\cal C}=  \frac{1}{\sqrt{g_{\text{CY}_2}}}\sum_i Q_i \delta(\underline{x}-\underline{x}_i)\,, \qquad  {\cal C}= \frac{L^2}{2 \lambda^2}(c_a)^2.
\eeq
We shall derive some solutions that follow from this in the next sections, starting with global $\mathbb{R}^4$.

\subsection{Global $\mathbb{R}^4$}
\label{sec:globalR4}

For the case of CY$_2=\mathbb{R}^4$ it is possible to construct solutions generalising that of section \ref{sec:basicsol}. For these we should refine \eqref{eq:CYsolgen} by first fixing
\beq
\label{Aleftforms}
ds^2(\text{CY}_2)= dr^2+\frac{r^2}{4}(L_a)^2\,, \qquad {\cal A}=\frac{r^2}{4}c_a L_a\,, \qquad dc_a=0\,.
\eeq
We shall refer to two unit radius 3-spheres in what follows: S$^3_1$ spanned by $(\psi,\text{S}^2)$ and S$^3_2$ spanned by $L_a$ -- they are the analogues of \eqref{eq:spheres}. If we assume a warp factor that respects the isometries of S$^3_2$, \eqref{eq:PDElocalR4} is solved in general in terms of three constants $(a_1,a_2,a_3)$ as
\beq
\label{fibredwarp}
h_5=a_1+\frac{a_2}{r^2}-a_3 r^2\,, \qquad  a_3=\frac{1}{8}\,{\cal C}\,.
\eeq
Generically such solutions will preserve an SU(2)$\times$ SU(2)$\times$U(1) isometry, though one can arrange for an additional U(1) by tuning $c_1=c_2=0$ (in terms of equation~(\ref{Fansatz}) this is equivalent to $f_1=f_2=0$). If we fix $a_3=0$ we turn off the fibration and recover the $\mathbb{R}^4$ solution of section \ref{sec:basicsol}, while for $a_3\neq 0$ the warp factor admits four zeros of the form
\beq
r_0= \pm \frac{a_1+s \sqrt{a_1+4 a_2 a_3}}{\sqrt{2 a_3}}\,, \qquad  s^2=1,
\eeq
each producing O5 plane behavior, thus depending on how we tune $(a_1,a_2,a_3)$ for $a_3\neq 0$ solutions can be bounded either between a D5 and O5 singularity or between two O5 planes -- in each case it is helpful to redefine the warp factor as
\beq
\label{h5D5O5O5}
h_5 \ = \  \left\{\begin{array}{l} \ds \frac{{\cal C}}{8}\frac{(b_1^2+r^2)(b_2^2-r^2)}{r^2}\,, \qquad  b_1^2>0\,, \qquad 0<r<b_2,\qquad \text{D5-O5}\\[2mm]
 \ds \frac{{\cal C}}{8}\frac{(b_1^2-r^2)(r^2-b_2^2)}{r^2}\,, \qquad  0<b_1<r<b_2\,, \qquad \text{O5-O5}
      \end{array}\right.
\eeq
Notice that there is also an intermediate case when $b_1=0$, where solutions now interpolate between a regular zero at $r=0$ and an O5 plane at $r=b_2$.

Let us consider these cases separately as they are physically distinct.

\subsubsection{D5-O5 case}
\label{sec:D5O5withF}

When we take the warp factor to be
\beq
\label{D5O5warp}
h_5= \frac{{\cal C}}{8}\frac{(b_1^2+r^2)(b_2^2-r^2)}{r^2}\,, \qquad b_2>0\,,
\eeq
the interval ${\cal I}$ spanned by $r$ is restricted to $[0,~b_2]$. At the lower bound of this interval the solution behaves as D5 branes wrapped on AdS$_3\times $S$^3_1$, and at the upper bound an O5 plane wrapped on this manifold. This behavior is similar to that of section \ref{sec:basicsol}, but we stress that now ${\cal F}$ is non trivial, so for this solution S$^3_1$ is non trivially fibered over S$^3_2$ and the 3-form flux has additional components turned on.  Despite the more complicated flux, the cycles over which we can define charges are unchanged and flux quantisation again requires that we impose that the quantities defined in \eqref{eq:R4fluxquantisation} are integers, this time we can achieve this by tuning the following combinations to be integer,
\beq
\label{chargesD5O5}
c=N_5\,, \qquad  \frac{c\lambda^4 b_2^4 (3 b_1^2+b_2^2){\cal C}}{768\pi^2} \ = \ N_1 \,, \qquad  \frac{c(b_1 b_2  \lambda)^2 {\cal C}}{8L^2}\ =\ n\,.
\eeq
given this we find the holographic central charge  via the formula \eqref{eq:centralcharge}
\beq
c_{hol}=  6 N_1 N_5\,,
\eeq
yielding the expected relation. As the sources are the same as those in section \ref{sec:basicsol}, it is tempting to again view  CY$_2$ as an orbifolded 4-sphere, this time however, the  S$^3_1$ would be fibered over the 4-sphere. Additionally due to the appearance of a non trivial ${\cal F}$ in \eqref{eq:sourcedPDE}, its integrated form imposes $\frac{(2\pi)^2}{8}b_2^4{\cal C}= \sum_i Q_i$ and as the LHS is strictly positive so is the RHS.  D5 branes contribute negatively to $\sum_i Q_i$, and as we have two O5s in 4-sphere interpretation, this time we can have only 1 D5 coincident to the O5  at $r=0$ rather than the 2 present when the fibration was trivial - hence $\frac{1}{8}b_2^4{\cal C}=1$. Of course it is possible that the internal space should be instead interpreted as a non compact space that is simply bounded, in which case there need be no such restriction on the D5 brane charge.

\subsubsection{O5-O5 case}
\label{sec:O5O5}

When we take the warp factor to be
\beq\label{eq:O5O5}
h_5 \ = \ \frac{{\cal C}}{8}\frac{(b_1^2-r^2)(r^2-b_2^2)}{r^2}\,,\qquad  0<b_1<b_2\,,
\eeq
the interval ${\cal I}$ spanned by $r$ is restricted to $[b_1,~b_2]$. This time one can show that the behavior close to both end points is that of an O5 plane wrapped on AdS$_3\times$S$^3_1$. The analysis runs  parallel to the previous example, with flux quantisation again demanding that we impose that
\eqref{eq:R4fluxquantisation} give integers, this time this means we should tune
\beq
\label{chargesO5O5}
c=N_5\,, \qquad \frac{c\lambda^4 (b_2^2-b_1^2)^3{\cal C}}{768\pi^2} \ = \ N_1 \,, \qquad \frac{c(b_1 b_2  \lambda)^2 {\cal C}}{ 8 L^2}=n\,,
\eeq
which again leads to a central charge of  $c_{hol}=  6 N_1 N_5$. If we are to again interpret this in terms of the orbifolded 4-sphere we no longer have any D5 branes present to compensate the O5 charge, however as the connection is non trivial this can be compensated for with its field strength. There are 2 units of O5 charge hence the integrated form of \eqref{eq:PDElocalR4} demands that we tune $\frac{1}{8}(b_2^4-b_1^4){\cal C}=2$. The intermediate case, bounded between an O5 plane and regular zero, behaves analogously -  one must tune $b_1=n=0$ and as there is now only 1 unit of O5 charge integrating \eqref{eq:PDElocalR4}  yeilds $ \frac{1}{8}b_2^4{\cal C}=1$.

\subsection{4-Torus}
\label{sec:4torus}

In the case of $\mathbb{T}^4$ we should refine \eqref{eq:CYsolgen} as
\beq
ds^2(\text{CY}_2)= \sum_{i=1}^4 (dx_i)^2,\qquad {\cal A}=c_1(x_{[1} dx_{2]}-x_{[3} dx_{4]})+  c_2 (x_{[3} dx_{1]}-x_{[2} dx_{4]})+c_3 (x_{[2} dx_{3]}-x_{[1} dx_{4]}),
\eeq
where one can take the coordinates to be periodic on the interval $x_i\in[0,R]$. The PDE one needs to solve is then 
\beq\label{eqT4genlap}
\sum_{i=1}^4\partial_{x_i}^2 h_5+ {\cal C}= \sum_k Q_k \delta(\underline{x}-\underline{x}_k)\,, \qquad {\cal C} = \frac{L^2}{2 \lambda^2}(c_a)^2.
\eeq
This type of  inhomogeneous generalised Laplace (Poisson) equation on the torus was considered in \cite{Andriot:2019hay,Andriot:2021gwv,Legramandi:2019ulq}. Note that as $\mathbb{T}^4$ is compact, integrating the PDE over it in general leads to the constraint 
\beq
{\cal C}= \frac{1}{\text{Vol}(\mathbb{T}^4)}\sum_i Q_i.
\eeq
The most simple scenario one can consider is to place a single source at $x_i= 0$, ie \eqref{eqT4genlap} becomes $\partial_{x_i}^2 h_5+ {\cal C}= Q \delta(\underline{x})$. Following \cite{Legramandi:2019ulq} one can naively deal with this  via a circular delta function of the form
\beq
\delta_R(x)=\frac{1}{R} \sum_{k=-\infty}^{\infty}e^{\frac{2\pi  i k x}{R}}
\eeq
such that $\delta(\underline{x})= \delta_R(x_1)\delta_R(x_2)\delta_R(x_3)\delta_R(x_4)$. Integrating the single source PDE over $\mathbb{T}^4$ then  yields $R^4{\cal C}=Q$, making $Q>0$ and so consistent with an O5 source and fixes ${\cal C}$ such that it precisely cancels the $O(1)$ term in $\delta(\underline{x})$. Taking a separations of variables ansatz for $h_5$ then leads to the solution
\beq
h_5= h_0- \frac{1}{R^2}\sum_{\underline{k}\in\mathbb{Z}^4/\underline{0}}\frac{1}{|\underline{k}|^2}e^{\frac{2\pi  i \underline{k}.\underline{x}}{R}}
\eeq
where $h_0$ is a constant and we have used that $Q^{\text{O5}}=(2\pi)^2$. As previously stated, this is the naive solution -- the series is not absolutely convergent. A more rigorous treatment can be found in \cite{Andriot:2019hay}, where it is shown that the general solution to a PDE of the form \eqref{eqT4genlap} is given in terms of Jacobi theta functions.


\subsection{Classes with CY$_2$ a circle fibration }
\label{sec:circlefibration}
A natural question to ask is what other non compact CY$_2$ manifolds can give rise to bounded solutions in 10 dimensions of the type in section \ref{sec:backgrounds}. Answering this question  requires details of the specific 4 manifold, so let us assume here that CY$_2$ contains  a U(1) isometry. The reason to do this, as we show in appendix \ref{sec:gency2withu1appendix},  is that all such manifolds fall into one of two classes both governed by a single PDE.\footnote{This result appears in several different contexts in the literature, see for instance \cite{Bakas:1996gf} and references there in.}\\
~~\\
\textbf{Class 1: U(1) fibration over $\mathbb{R}^3$}\\
~~\\
The first class of CY$_2$ solutions containing a U(1) isometry $\partial_{\phi}$ can be expressed locally as
\beq
ds^2(\text{CY}_2)= \frac{1}{\partial_y H}(d\phi-\partial_{x_2}H dx_1+\partial_{x_1}H dx_2)^2+ \partial_y H\left(dy^2+dx_1^2+dx_2^2\right),
\eeq
as such these solutions are U(1) fibrations over $\mathbb{R}^3$, where $H$ must obey a Laplace equation on the $\mathbb{R}^3$ spanned by $(x_1,x_2,y)$
\beq
\nabla^2_{\mathbb{R}^3} H=0,
\eeq
where we stress that this equation is exact, it cannot have source terms as that would make the manifold not CY.
For this class, the corresponding Killing spinors on CY$_2$ are $\partial_{\phi}$ singlets.\\
~~\\
To embed such manifolds into the class of section \ref{sec:backgrounds} with non trivial fibration, we first need to define ${\cal F}$, this can be taken to be as in \eqref{Fansatz} for the canonical vielbein of \eqref{eq:complexvielbein}.\footnote{More specifically, one should decompose $E_1= e^1+i e^2,~E_2= e^3+i e^4$ then fix the functions of the vielbein as appendix \ref{sec:gency2withu1appendix} describes.} If we take $\partial_{\phi}$ to be an isometry of ${\cal F}$, then that it should also be closed requires
\beq
f_a=\partial_{z_a}G,~~~~ 2\partial_{z_a}G\partial_{z_a}\log(\partial_{y}H)+ \nabla^2_{\mathbb{R}^3}G=0,
\eeq  
where $z_a=(x_1,x_2,y)$ and $G$ is independent of $\phi$. To have a solution one must also solve the following ODE for the warp factor
\beq\label{eq:cycirfib1}
\nabla_{\mathbb{R}^3}^2 h_5+\frac{L^2\partial_y H}{2\lambda^2}(\partial_{z_i}G)^2=0,
\eeq
away from the loci of sources, we have assumed $\partial_{\phi}$ is also an isometry of $h_5$.  The simplest solution to this system is to take $h$ and $G$ to be linear functions of $y$ only, one then has locally that CY$_2= \mathbb{T}^4$ and, generalising \eqref{eq:cycirfib1} to include source terms, we reproduce \eqref{eqT4genlap} of the previous section, specialised to the case where one direction in $\mathbb{T}^4$ remains an isometry in the full space.
\\
~\\
\textbf{Class 2: U(1) fibrations governed by a Toda equation}\\
~~\\
The second class of U(1) preserving  CY$_2$ manifolds is the more interesting, these can locally be expressed as
\beq
ds^2(\text{CY}_2)=\frac{1}{\partial_y\Delta}\left(d\phi-\partial_{x_2}\Delta dx_1+\partial_{x_1}\Delta dx_2\right)^2+\partial_y\Delta\left(dy^2+ e^{2\Delta}(dx_1^2+dx_2^2)\right),
\eeq
where $\Delta$ is governed by the 3 dimensional Toda equation
\beq\label{eq:toda}
2\nabla_2^2 \Delta+ \partial_{y}^2 e^{2\Delta}=0,
\eeq
where $\nabla_2^2=\partial_{x_1}^2+\partial_{x_2}^2$. Note that this PDE also appears in the context of ${\cal N}=2$ AdS$_5$ solutions \cite{Lin:2004nb}, albeit here, like the previous example it should not have source terms. For this class of solutions the 3 manifold over which the U(1) is fibered itself depends on how the Toda equation is solved and the corresponding Killing spinor is charged under $\partial_{\phi}$.\\
~\\
To embed such manifolds into  section \ref{sec:backgrounds} we again take the ansatz of \eqref{Fansatz} for the vielbein of \eqref{eq:complexvielbein} and impose that $\partial_{\phi}$ is an isometry of ${\cal F}$. Imposing $d{\cal F}=0$ then requires that we fix
\beq
f_1= \partial_y g,~~~~f_2=e^{-\Delta}\partial_{x_1}g,~~~~f_3=e^{-\Delta}\partial_{x_2}g,~~~~ g=g(y,x_1,x_2),
\eeq
and that we solve the PDE
\beq\label{eq:nottoda}
2\big(\partial_{x_1}g\partial_{x_1}(\log\partial_y\Delta)+\partial_{x_2}g\partial_{x_2}(\log\partial_y\Delta)\big)+\frac{\partial_yg}{\partial_y\Delta}\partial_y^2(e^{2\Delta})+\nabla_2^2g+e^{2\Delta}\partial^2_y g=0.
\eeq
The PDE governing the warp factor $h_5$, assuming it too is independent of $\phi$, is then
\beq\label{eq:h5toda}
\nabla_2 h_5+\partial_y(e^{2\Delta} \partial_y(h_5))+\frac{L^2\partial_y \Delta}{2\lambda^2}\big((\partial_{x_1} g)^2+(\partial_{x_2} g)^2+e^{2\Delta}(\partial_yg)^2\big)=0\,,
\eeq
away from the loci of sources.

An example of a solution in this class is the local solution of section \ref{sec:globalR4}, this is given by fixing 
\beq
e^{2\Delta}= \frac{4y^2}{(1+x_1^2+x_2^2)^2},~~~~g=-\frac{2y}{1+x_1^2+x_2^2}\big(c_1 x_1+c_2 x_2+\frac{c_3}{2}(1-x_1^2-x_2^2)\big)\,,
\eeq
so that $(x_1,x_2)$ span a 2-sphere in stereographic coordinates. This solves both \eqref{eq:toda} and \eqref{eq:nottoda}, \eqref{eq:h5toda} then yields \eqref{fibredwarp} if one assumes $h_5=h_5(y)$ and identifies $4y=r^2$.\\
~~\\
It is our hope that the content of this section will lead to the construction of more broad classes of generalised D1-D5 near horizon solutions, ie for CY that are not locally $\mathbb{R}^4$, but that is beyond the scope of this work.

\section{A consistent truncation to minimal $d=5$  supergravity coupled to an Abelian vector multiplet}
\label{sec:Sdual}
It is well known that the D1-D5 near horizon admits a consistent truncation (on CY$_2$) to minimal $d=6$ (ungauged) supergravity, indeed AdS$_3\times$S$^3$ is one of just 3 maximally supersymetric vacua of this theory \cite{Gutowski:2003rg}. It is shown in \cite{Andrianopoli:2004xu} (see also \cite{Hristov:2014eba}) that the $d=6$ theory can itself be consistently truncated to minimal $d=5$ supergravity coupled to an abelian multiplet, which AdS$_3\times$S$^2$ is vacua of. As the class of \eqref{eq:CYsolgen} is generically ${\cal N}=(4,0)$ with the superconformal algebra realised by the AdS$_3$ and S$^2$ factors in the metric, it is then reasonable to ask whether this too admits a consistent truncation to $d=5$, in this section we establish this is indeed the case\footnote{For resent related work on consistent truncations to $d=5,6$ gauged and un-gagued supergravities see \cite{Couzens:2022aki,Lozano:2022ouq}}.\\
~~\\
The most general form of $d=5$ supergravity in 5 dimensions is presented for instance in \cite{Behrndt:1998ns} (the gauge free case is when $g=0$). Its bosonic sectors consists of the metric, $n$ gauge fields $A^I$ with abelian fields strengths $G^I$ and $n$ scalar fields $X^I$ subject to the constraint
\beq
{\cal V}=C_{IJK}X^IX^JX^K=1,~~~~ I,J,K=1,...,n
\eeq
with specific models defined via a choice of ${\cal V}$.  We shall concern ourselves with a specific ungauged model with $n=2$ and
\beq
C_{122}=C_{212}=C_{221}=2,~~~~ C_{IJK}=0 ~~~~\text{otherwise}, ~~~~~ (X^1)^2=(X^2)^{-1}= e^{2\phi}.
\eeq
considered  in \cite{Hristov:2014eba}, in the context of reductions between 6 and 4 dimensions. The action of the bosonic sector of this model takes the form
\beq
S^{(5)}=\int d^5x\sqrt{-g^{(5)}}\bigg(\frac{1}{2} R^{(5)} -\frac{3}{8}(\nabla^{(5)}\phi)^2-\frac{1}{8}e^{-2\phi}G_1^2-\frac{1}{4}e^{\phi}G_2^2\bigg)-\frac{1}{12}\int C_{IJK}G_I\wedge G_J\wedge A_K.\label{eq:action5d}
\eeq
These fields form a gravity and vector multiplet, $(g^{(5)}_{\alpha\beta},A^{\alpha}_1)$ and $(A^{\alpha}_2,e^{\phi})$ respectively. The conditions for unbroken supersymmetry are presented in \cite{Behrndt:1998ns}, 8 real supercharges is maximal and the reduction of the D1-D5 near horizon  is $\frac{1}{2}$ BPS \cite{Hristov:2014eba}.\\
~~\\
To derive a consistent truncation we must generalise \eqref{eq:CYsolgen} such that it contains the same field content as \eqref{eq:action5d}. To this end we make the ansatz\footnote{So as not to make the presentation overly long, we are simply presenting the answer, to actually derive this one needs to start with a more general ansatz depending on only the bosonic fields of the $d=5$ theory.  For instance for metric one can take
\beq
ds^2= \frac{L^2}{\sqrt{h_5}}\bigg[ e^{c_1\phi}g^{(5)}_{\mu\nu} dx^{\mu}dx^{\mu}+ \frac{1}{4}e^{c_2\phi}{\cal D}\psi^2\bigg]+ \lambda^2e^{c_3\phi}\sqrt{h_5}ds^2(\text{CY}_2),~~~~{\cal D}\psi= d\psi+ c_4 A_1+ c_5 A_2+{\cal A}\nn
\eeq
where $c_i$ are arbitrary constants. One then takes similar ansatze for the dilaton and 3-form, depending on more arbitrary constants. Finally one fixes these constants by demanding the $d=10$ equations of motion hold when the $d=5$ equations of motion, \eqref{eq:unsourcedPDE} and the properties of ${\cal F}$ (closed and anti-selfdual) are assumed to.}
\begin{align}
ds^2&= \frac{L^2}{\sqrt{h_5}}\bigg[ e^{-\frac{1}{4}\phi}g^{(5)}_{\mu\nu} dx^{\mu}dx^{\mu}+ \frac{1}{4}e^{\frac{3}{4}\phi}{\cal D}\psi^2\bigg]+ \lambda^2e^{\frac{3}{4}\phi}\sqrt{h_5}ds^2(\text{CY}_2),~~~~ e^{\Phi}=e^{\frac{3}{4}\phi}\frac{L^2}{ c\sqrt{h_5}},\nn\\[2mm]
F_3&= c e^{-2\phi}\star_5 G_1+ \frac{c\lambda^2}{L^2}\star_4 dh_5+\frac{c}{4}{\cal D}\psi\wedge \bigg[ 2 G_2- {\cal F}\bigg],~~~~{\cal D}\psi= d\psi+ 2A_2+{\cal A}.\label{eq:NSred}
\end{align}
One then needs to check that the ansatz solves the type IIB equations of motion. In string frame, these can be expressed in the form
\begin{subequations}
\begin{align}
&dF_3=0,~~~~ d\star F_3=0,\label{eq10deom1}\\[2mm]
&{\cal D}=e^{2\Phi}(\nabla^2\Phi-2 (\nabla \Phi)^2)-\frac{1}{2} F_3^2=0,\label{eq10deom2}\\[2mm]
&{\cal E}_{AB}=e^{-2\Phi}\left(R_{AB}+2\nabla_{A}\nabla_B\Phi+\frac{1}{4}g_{AB}(\nabla^2\Phi-2 (\nabla \Phi)^2)\right)-\frac{1}{4} (F_3^2)_{AB}+\frac{1}{48}g_{AB}F_3^2=0,\label{eq10deom3}
\end{align}
\end{subequations}
where $(A,B,...)$ are ten dimensional indices and we define $F_3^2= (F_3)_{ABC}(F_3)^{ABC}$ and $(F_3^2)_{AB} =(F_3)_{A}^{~~CD}(F_3)_{BCD}$. It is a simple matter to show that the Bianchi identity and EOM of $F_3$ \eqref{eq10deom1}  reduce to
\beq
d(e^{-2\phi}\star_5 G_1)+G_2\wedge G_2 =0,~~~~~ d(e^{\phi}\star_5 G_2)+G_1\wedge G_2=0,
\eeq    
given \eqref{eq:unsourcedPDE}, that $(G_1,G_2,{\cal F})$ are closed and ${\cal F}$ is anti self dual.
It is not hard to show that this is precisely what one gets by varying \eqref{eq:action5d} with respect to $A_1$ and $A_2$ respectively. The ten dimensional dilaton EOM \eqref{eq10deom2} expanded out on our ansatz becomes 
\beq
{\cal D}= \frac{c^2e^{-\frac{9}{4}\phi}}{2L^4 \lambda^2 \sqrt{h_5}}\left(\nabla^2_{\text{CY}_2}h_4+\frac{L^2}{8\lambda^2}{\cal F}^2\right)-\frac{c^2}{L^6 \sqrt{-g^{(5)}}} e^{-\frac{5}{4}\phi}h_5^{\frac{3}{2}}\frac{\delta S^{(5)}}{ \delta \phi},
\eeq
with the first term vanishing by \eqref{eq:unsourcedPDE} and the second by the EOM of the $d=5$ theory. Proving that all components of ${\cal E}_{AB}=0$ is a more lengthy computation, though this also follows from the EOM that \eqref{eq:action5d} implies, \eqref{eq:unsourcedPDE}, the anti-self duality of ${\cal F}$ and
\beq
{\cal F}_{ac}{\cal F}_{b}^{~c}\ = \ \frac{1}{4}\,g^{(4)}_{ab}{\cal F}^2\,,
\eeq
where $a,b...$ are indices on CY$_2$ and $g^{(4)}_{ab}$ is its unwarped metric. This final conditions actually follows from the anti-self duality of ${\cal F}$. Finally we should stress that, although we have not spoken of them explicitly in this section until now,  there is no barrier to performing this truncation in the presence of sources. If they are consistent for \eqref{eq:CYsolgen}, then they are too for \eqref{eq:NSred} as they  cancel between the $D=10$ EOM and the generalised Laplace for $h_5$ in the same way for each.\\
~~\\
We have thus derived a new embedding of $d=5$ ungauged supergravity+ abelian vector multiplet into type IIB supergravity, containing a circle fibration over CY$_2$ and potential source terms. When a solution to the $D=5$ supergravity is supersymmetric, we expect the lifted solution to also preserve supersymmmetry, though we have not checked this explicitly.




\section{Spin 2 mode}
\label{sec:spin2}

In this and the next section we will consider two examples of linear perturbations of type IIB supergravity about backgrounds of section~\ref{sec:backgrounds}. This will be a first step in the computation of the Kaluza-Klein spectrum and possibly the operator spectrum in the dual CFT$_2$. Computing the full spectrum is a difficult task in general, indeed until recently this had only been managed for solutions that decompose as direct products of co-set spaces (i.e. AdS$_4\times$S$^7$~\cite{Biran:1983iy}, AdS$_5\times$S$^5$~\cite{Kim:1985ez}, AdS$_5\times$T$^{1,1}$~\cite{Ceresole:1999zs}, AdS$_3\times $S$^3\times $K3~\cite{Deger:1998nm}, see however~\cite{Berg:2006xy,Benna:2007mb,Dymarsky:2008wd,Gordeli:2009nw,Melnikov:2020cxe} for the detailed analysis of a subsector of a warped AdS$_5\times$T$^{1,1}$). This changed with the advent of so-called ``Kaluza-Klein spectrometry''~\cite{Malek:2019eaz,Malek:2020yue,Eloy:2020uix,Bobev:2020lsk,Cesaro:2021haf}, which uses the frame work of exceptional field theory to compute the spectrum of any solution admitting a consistent truncation to a maximal gauged supergravity. Unfortunately this frame work is rather less developed for AdS$_3$ vacua than their higher dimensional counter parts, see~\cite{Eloy:2020uix} for an example. Additionally spectroscopy for solutions that do not admit a truncation to maximal or half maximal (for the case of AdS$_3$) gauged supergravity is yet to be worked out. The solutions within section~\ref{sec:backgrounds} hit both of these stumbling blocks, so in this work we shall set ourselves the more modest goal of finding some modes that decouple from the main system, allowing us to compute their spectra with more traditional means. 

In \cite{Bachas:2011xa} a universal equation was derived for spin 2 fluctuations on maximally symmetric spaces embedded into type IIB supergravity\footnote{One of the early papers noting this universality of spin-two fluctuation is~\cite{Csaki:2000fc}.}. The original derivation is done for 4-dimensional symmetric spaces, but it is straightforward to generalise it to the general case. The starting point is a metric of the form
\be
ds^2= G_{MN} dX^{M}dX^N \ =\  e^{2A}\overline{g}_{\mu\nu}dx^{\mu}dx^{\nu}+ \hat{g}_{ab}dy^a dy^b\,.
\ee
One considers the fluctuations along the symmetric space directions $x^\mu$,
\beq
\label{spin2fluc}
\delta g_{\mu\nu}\ = \ e^{2A}h_{\mu\nu}\,, \qquad  \delta \hat g_{ab}\ =\ 0\,.
\eeq
In the Einstein frame $h_{\mu\nu}$ satisfies the $d$-dimensional Laplace-Beltrami equation,
\beq
\frac{1}{\sqrt{G}}\partial_M\left(\sqrt{G}G^{MN}\partial_N\right)h_{\mu\nu} \ = \ 0\,.
\eeq
After separating $x$ and $y$ coordinates,
\beq
\label{IntSpaceFunction}
h_{\mu\nu} \ = \ h^{tt}_{\mu\nu}(x) \Psi(y)\,,
\eeq
and imposing transverse-traceless condition (with respect to the symmetric space metric $\overline{g}_{\mu\nu}$),  
\beq
 \overline{g}^{\mu\nu}h^{tt}_{\mu\nu} \ = \  \overline{\nabla}^{\mu}h^{tt}_{\mu\nu} \ = \ 0,
\eeq
one can separate the symmetric space part of the linearized Einstein equations, 
\beq\label{eq:spin2eq}
\overline{\nabla}^2 h^{tt}_{\mu\nu}\ =\ (M^2+ 2k) h^{tt}_{\mu\nu}\,,
\eeq
where $k$ is the scalar curvature of the symmetric space: $k=-1,0,1$ for anti de Sitter, Minkowski and de Sitter spaces respectively. This equation is equivalent to the equation of a scalar particle of mass $M$ in the symmetric space. From the point of view of the AdS/CFT correspondence ($M$ being AdS$_n$ space) the equation describes an operator of dimension
\be
\Delta \ = \ \frac{n-1}{2} +\sqrt{\left(\frac{n-1}{2}\right)^2+ M^2}
\ee
in the dual CFT$_{n-1}$. Here $M$ and $k$ are assumed to be in units of the curvature radius, whenever appropriate.

As in standard separation of variables, the mass $M$ is determined by an eigenvalue problem for the second factor $\Psi(y)$ in (\ref{IntSpaceFunction}), in the internal space. For an $n$-dimensional maximally symmetric space embedded into $d$ total dimensions the generalisation of the equation for $\Psi(y)$ is
\beq
\bigg[\frac{e^{(2-n)A}}{\sqrt{\hat g}}\partial_{y_a}\big(e^{n A}\sqrt{\hat g}\hat g^{ab}\partial_{y_b}\big)\bigg]\Psi \ = \ - M^2 \Psi \,.
\label{eq:spin2formula}
\eeq
The warp factor deforms the standard Laplacian on the internal space. We remind the reader that this is an Einstein frame expression, to convert to string frame we use the identity  $ds^2_E= e^{-\frac{\Phi}{2}} ds^2_S$ so that
\beq
e^{2A} \ = \ e^{2A_S-\frac{\Phi}{2}}\,,\qquad \hat g_{ab} \ = \  e^{-\frac{\Phi}{2}}\hat g^S_{ab}\,, \qquad \sqrt{\hat g} \ =\  e^{-\frac{(d-n) \Phi}{4}}\sqrt{\hat g^S}\,, 
\eeq
note $d=10$, and then
\beq
\label{spin2onCompact}
\bigg[\frac{e^{(2-n)A_S+2\Phi}}{\sqrt{\hat g^S}}\partial_{y_a}\big(e^{n A_S-2\Phi}\sqrt{\hat g^S} (\hat g^S)^{ab}\partial_{y_b}\big)\bigg]\Psi\ = \ - M^2 \Psi.
\eeq
For the background~(\ref{eq:CYsolgen}) with~${\cal{F}}=0$, as the one of equation~(\ref{eq:solsimp}) in section~\ref{sec:basicsol}, we have $n=3$, and $k=-1$. In this case the internal space equation takes the form
\beq\label{eq:spin2eigenfunctioneq}
\left(\nabla_{\text{S}^3}^2+ \frac{L^2}{h_5 \lambda^2} \nabla_{\text{CY}_2}^2\right)\Psi \ = \  - M^2 \Psi\,.
\eeq
In particular, for the zero modes on the CY$_2$ the effect of the warping is non-existent. For the fibred solutions~(\ref{eq:CYsolgen}) with ${\cal F}\neq 0$, i.e. the ones considered in section~\ref{sec:fibredsols}, the equation becomes
\beq
\label{spin2fibred}
\bigg(\nabla_{\text{S}^3}^2+ \frac{L^2}{h_5 \lambda^2} \bigg[\nabla_{\text{CY}_2}^2+ |{\cal A}|^2\partial_{\psi}^2-2 \nabla_{{\cal A}} \partial_{\psi}-(\nabla_{\text{CY}_2}. {\cal A})\partial_{\psi}\bigg]\bigg)\Psi\ = \  - M^2 \Psi\,,
\eeq 
where $|{\cal A}|^2$ is the norm defined on CY$_2$ rather than its warped equivalent, similarly for $\nabla_{{\cal A}}$ and one is free to choose a gauge in which $\nabla_{\text{CY}_2}. {\cal A}=0$.

One can also show that this perturbation of the metric decouples from all the remaining fluctuations of the type IIB fields and the D brane sources. Consequently, equation~(\ref{eq:spin2formula}) describes a consistent spin 2 mode for a large number of holographic backgrounds. In the examples below we will analyse the spectrum of the dual operators described by this mode.

\subsection{Large ${\cal{N}}=(4,0)$ example: AdS$_3\times$S$^3\times $S$^3 \times \mathbb{R}$}
\label{sec:simplespec}

Before discussing the spectrum on the backgrounds described in sections~\ref{sec:backgrounds} and~\ref{sec:fibredsols} we would like to consider a more simple class of solutions. One of the purposes of this exercise is to see the interplay of the BPS conditions in supergravity and CFT. We will also use the simple example to introduce general boundary conditions to be used in the remaining part.

In \cite{Macpherson:2018mif} a solution with D8 branes on  AdS$_3\times$S$^3\times $S$^3 \times \mathbb{R}$ was constructed. For this one takes
\beq
\label{D8O8metric}
e^{2A} \ = \ \frac{L^2}{\sqrt{h_8}}\,, \qquad e^{-\Phi} \ = \ c \, h_8^{\frac{5}{4}}\,, \qquad  ds^2(\text{M}_7) \ = \ c^2\sqrt{h_8} dr^2+\frac{L^2}{\sqrt{h_8}}\bigg[\frac{1}{\cos^2\!\beta}ds^2(\text{S}^3_1)+\frac{1}{\sin^2\!\beta}ds^2(\text{S}^3_2)\bigg]\,,
\eeq
where $L$, $c$ and $\beta$ are constants. The parameter $\beta$ sets the relative radii of the spheres in such a way that
\beq
\frac{1}{R_{\text{S}_1^3}^2} + \frac{1}{R_{\text{S}_2^3}^2} \ = \ \frac{1}{L^2} \ = \ \frac{1}{R_{\text{AdS}_3}^2}\,.
\eeq
The warp factor $h_8$ is a piecewise linear function such that in a linear interval the D8 brane flux is given by
\beq
\partial_r h_8 \ = \ F_0\,,
\eeq
where $F_0$ is the value of the flux that jumps at the position of D8 branes along $r$ direction producing a discontinuity in $\partial_r h_8$. A compact solution is an interval bounded between two D8/O8 systems, say at $r=r_\pm$. Since $h_8(r_\pm)=0$ there should be at least one locus of D8 branes inside the interval. Patches of continuous solutions should be glued appropriately at the loci of D8.

From \eqref{eq:spin2formula}, we find that in the background~(\ref{D8O8metric}),
\beq
\bigg[\frac{L^2}{c^2 h_8}\partial_r^2 + \cos^2\!\beta\,\nabla^2_{S^3_1}+\sin^2\!\beta\,\nabla^2_{S^3_2}+M^2\bigg]\Psi \ = \ 0\,.
\eeq
This has a structure similar to~(\ref{eq:spin2eigenfunctioneq}). Taking the mode expansion
\beq
\label{harmexp}
\Psi\ = \  \sum_{\ell_1,\ell_2} H_{\ell_1,\ell_2}(r) Y^{\ell_1}_{\text{S}^3_1}Y^{\ell_2}_{\text{S}^3_2}\,,
\eeq
where the scalar harmonics on the spheres satisfy $\nabla^2_{S^3_i}Y^{\ell_i}_{\text{S}^3_i}=(1-(\ell_i+1)^2)Y^{\ell_i}_{\text{S}^3_1}$. Since the equation has SO(4)$\times$SO(4) symmetry we suppress the summation over the projections of the spins, which label the complete eigenfunction basis (see Appendix~\ref{sec:harmonics} for more details). We find
\beq
\label{StokesEq}
\frac{1}{h_8}\partial^2_r H_{\ell_1\,\ell_2}+ \frac{\kappa^2_{\ell_1,\ell_2}c^2}{L^2} H_{\ell_1,\ell_2} \ =\ 0\,, \qquad  \kappa^2_{\ell_1,\ell_2} \ =\ M^2- \ell_1(\ell_1+2)\cos^2\!\beta-\ell_2(\ell_2+2)\sin^2\!\beta\,.
\eeq
Consequently one needs to solve a Sturm-Liouville problem for this equation, for which a choice of boundary conditions should be made on the interval $r\in [r_-,r_+]$. Let us first discuss boundary conditions at the endpoints (O8 positions). One way to motivate the choice is to note that the variational principle for fluctuations requires a treatment of the boundary terms in the variation of the action, cf.~\cite{Bachas:2011xa}. From the form of equation~(\ref{spin2onCompact}) one can see that the universal boundary terms are of the form
\be
\delta \Psi\, e^{3A - 2\Phi}\sqrt{\hat g} \hat{g}^{rr}\partial_{r}\Psi\bigg|_{r=r_\pm} \ \to \ 
\delta H_{\ell_1,\ell_2}\partial_r H_{\ell_1,\ell_2}\bigg|_{r=r_\pm}\,.
\ee
For the variational principle to be completely defined one needs to project out these boundary terms by either imposing Dirichlet boundary conditions $\delta H_{\ell_1,\ell_2}|_{r=r_\pm}=0$, that is choosing fixed values for the fluctuations on the boundary, or Neumann boundary conditions $\partial_r H_{\ell_1,\ell_2}|_{r=r_\pm}=0$. Note that for generic supergravity fields the above boundary terms, and consequently boundary conditions can receive additional non-universal terms coming from D-brane sources, localized on the boundaries. However, it was argued in~\cite{Bachas:2011xa} that such boundary sources decouple from the spin two mode considered here.

Note that for the internal geometries of the form S$^3\times$CY$_2$  fibred over an interval $r\in[r_-,r_+]$, which we consider here, equation~(\ref{spin2onCompact}) has the canonical form of a Sturm-Liouville problem,
\be
\label{canonicalSL}
\partial_r\left(p(r)\partial_r H_{\ell_1\,\ell_2}\right) + \beta_{\ell_1,\ell_2}^2 q(r)H_{\ell_1\,\ell_2} \ = \ 0\,,
\ee
with
\be
\label{pandq}
p(r) \ = \ e^{3A - 2\Phi}\sqrt{\hat g} \hat{g}^{rr}\,, \qquad \text{and} \qquad q(r) \ = \ e^{A - 2\Phi}\sqrt{\hat g}\,,
\ee
modulo constants that can be absorbed into $M^2\to\beta^2_{\ell_1,\ell_2}$. Hence, the Neumann boundary conditions are generalized to 
\begin{eqnarray}
\label{Neumann}
\text{Neumann:} & p(r)\partial_rH_{\ell_1,\ell_2}\bigg|_{r=r_\pm} \ = \ 0\,,
\end{eqnarray}
allowing for a singularity of $\partial_r H_{\ell_1,\ell_2}$ at the boundary if $p(r)$ vanishes there. This boundary condition implies positivity of the eigenvalues $\beta_{\ell_1,\ell_2}^2$ (unitarity),
\begin{multline}
\label{NormPos}
\beta_{\ell_1,\ell_2}^2||H_{\ell_1,\ell_2}||^2 \ \equiv \ \beta_{\ell_1,\ell_2}^2\int\limits_{r_-}^{r_+} dr\  q(r)\left(H_{\ell_1,\ell_2}\right)^2 \ = \  \int\limits_{r_-}^{r_+} dr \  p(r)\left(\partial_r H_{\ell_1,\ell_2}\right)^2 \\ - \beta_{\ell_1,\ell_2}^2 p(r)H_{\ell_1,\ell_2}\partial_r H_{\ell_1,\ell_2}\bigg|_{r_-}^{r_+} \ \geq \ 0\,.
\end{multline}
Besides the boundary conditions, we require the natural norm defined with respect to the weight $q(r)$ to be finite.

For the Dirichlet boundary conditions to be consistent with the bound $\beta_{\ell_1,\ell_2}^2\geq 0$, one has to select a stronger form of the latter,\footnote{These boundary conditions are consistent with the ones advocated in~\cite{Passias:2016fkm,Passias:2018swc}, where a similar analysis was performed.}
\begin{eqnarray}
\label{Dirichlet}
\text{Dirichlet:} & H_{\ell_1,\ell_2}\bigg|_{r=r_\pm} \ = \ 0\,.
\end{eqnarray}
The final choice of the boundary conditions depends on further details of the problem. As outlined in the introduction discussion of the conditions at the loci of D branes and O planes is in general tricky, because of possible ambiguities and because the supergravity approximation breaks down in the vicinity of these objects. However, in some cases, a natural choice can be made based on the analyticity and normalizability of fluctuations.

In terms of the above problem, the non-negativity of the eigenvalues $\beta^2_{\ell_1,\ell_2}\geq 0$ implies
\beq
\label{LargeN4bound}
M^2 \ \geq \ \ell_1(\ell_1+2)\cos^2\!\beta+\ell_2(\ell_2+2)\sin^2\!\beta \,.
\eeq
This relation can be compared with the BPS bounds in the supergravity algebra D$(2,1|\alpha)$~\cite{Gunaydin:1988re,deBoer:1999gea}, or in the large ${\cal N}=(4,0)$ superconformal algebra $A_\gamma$ of the dual CFT~\cite{Sevrin:1988ew}. The three bounds agree if and only if $\ell_1=\ell_2=\ell$. Indeed, the analysis of the shortening condition in the D$(2,1|\alpha)$ superalgebra implies
\be
\label{D21alphaBPS}
\Delta_{(2)} = 1+\sqrt{1+M^2} \ \geq \ell_1\cos^2\!\beta +\ell_2\sin^2\!\beta + 2\,, \qquad \alpha = \tan^2\!\beta\,, \qquad \gamma \ = \ \sin^2\!\beta\,,
\ee
where the shift by two units appears because we compare the upper components of the multiplets, and the above statement follows. Moreover, one can also see that states with $\ell_1\neq \ell_2$ saturating the D$(2,1|\alpha)$ bound~(\ref{D21alphaBPS}) violate the eigenvalue bound~(\ref{LargeN4bound}) and therefore are absent from the supergravity spectrum\footnote{The fact that supergravity imposes the condition $\ell_1=\ell_2$ on the D$(2,1|\alpha)$ spectrum was noted in~\cite{Eberhardt:2017fsi} in the context of AdS$_3\times$S$^3\times$S$^3\times$S$^1$ background. This helped to explain the absence of the superfluous states from the early Kaluza-Klein analysis of~\cite{deBoer:1999gea}.} .

Finally, we note that only Neumann boundary conditions for $H_{\ell_1,\ell_2}$ are compatible with the BPS spectrum, since the zero modes of~(\ref{canonicalSL}) are simply constant functions. We note that Neumann boundary conditions at the locus of the O8 are compatible with the symmetry of the orbifold action, so we will focus on these boundary conditions for the O planes in the rest of this paper. Boundary conditions at the D-brane positions will depend on the specific setup. In particular, at the loci of D8 inside the interval, we demand continuity of the function and its derivative, since discontinuities would manifest themselves through delta functions, that is source terms in the equation, while the spin two equation was shown to decouple from any sources, cf. the discussion in section 4.3 of~\cite{Passias:2016fkm}.

As far as the non-BPS spectrum is concerned, we will consider several different cases. First, $h_8=1$ is the case of the trivial warping, for which one obtains
\beq
H_{\ell_1,\ell_2} \ = \ A_{\ell_1,\ell_2}\cos\left(\frac{\kappa_{\ell_1,\ell_2}c}{L}r\right)+ B_{\ell_1,\ell_2}\sin\left(\frac{\kappa_{\ell_1,\ell_2}c}{L}r\right).
\eeq
If the space is non-compact, the spectrum is continuous. If we instead consider the compact case AdS$_3\times$S$^3\times $S$^3 \times $S$^1$, there is a discrete tower of non-BPS states with
\beq
M^2 \ = \ \ell_1(\ell_1+2)\cos^2\!\beta+\ell_2(\ell_2+2)\sin^2\!\beta + \frac{4\pi^2n^2L^2}{(r_+-r_-)^2c^2}\,,
\eeq
labeled by integer $n$. 

Since in more general case the warp factor is a linear function, one can express the solution to equation~(\ref{StokesEq}) in terms of Airy functions, albeit of a complex argument. Let us consider the situation $r_\pm=\pm r_0$ and D8 branes at $r=0$, so that
\beq
h_8 = \left\{\begin{array}{ll}
F_0\cdot (r_0+r)\,,  & -r_0\leq r < 0\,,     \\
F_0\cdot (r_0-r)\,, & 0 < r \leq r_0      
\end{array} \right. 
\eeq
The general solutions to equation~(\ref{StokesEq}) satisfying Neumann boundary conditions $H_{\ell_1,\ell_2}'(\pm r_0)=0$ at $r=\pm r_0$ are
\beq
H_{\ell_1,\ell_2} \ = \ A_{\ell_1,\ell_2}^{\pm}\ _0F_1\left(\frac23;\frac{F_0\kappa_{\ell_1,\ell_2}^2}{9}\frac{c^2}{L^2}(\pm r-r_0)^3\right).
\eeq
Gluing together the $\pm$ branches at $r=0$ requires the full solution to be an even function. The problem reduces to finding zeroes of the derivative of the hypergeometric function at $r=0$. The corresponding discrete spectrum can be found numerically or via the WKB method, which gives the following approximation:
\be
M^2\simeq \ell_1(\ell_1+2)\cos^2\!\beta+\ell_2(\ell_2+2)\sin^2\!\beta + \frac{9\pi^2n^2L^2}{4F_0r_0^3c^2}\,.
\ee


\subsection{AdS$_3\times$S$^3\!\times\!\mathbb{R}^4$ background}
\label{sec:spin2noF}

Now we turn to the case of AdS$_3\times$S$^3\!\times\!\mathbb{R}^4$ geometry introduced by solution~(\ref{eq:solsimp}) in section~\ref{sec:basicsol}. We remind the reader that the geometry describes D5 branes and O5 planes back reacted on $\mathbb{R}^4$. As discussed in section~\ref{sec:basicsol}, one has an O5 located at $\rho=0$ and either a single D5, or a pair of D5 with an O5 at $\rho=\pi/2$\footnote{We start this section using coordinate $\rho$ (not to be mixed with the spacetime index), which is the convention of~\cite{Macpherson:2018mif}, where the background solution was originally constructed, and of section~\ref{sec:basicsol}. However, the equations are simpler when written in terms of the natural radial coordinate $r=\cos\rho$.}. Consequently, we shall refer to the above points as O5 and D5 loci respectively. 

Following the general discussion at the beginning of this section we consider traceless transverse metric perturbation along the AdS$_3$ directions\footnote{From this moment on, unless explicitly stated, the capital indices $M$, $N$, $P$, $\ldots$ are 10-dimensional indices, lowercase Greek indices label AdS$_3$ direction, lowercase Latin indices $a$, $b$, $c$ label the directions of the first S$^3$ factor and $i$, $j$, $k$ -- the directions along S$^3$ in CY$_2$.},
\be
\delta g_{\mu\nu} \ =  \ \cot \rho\, h_{\mu\nu}\,, \qquad h_{\mu\nu}g^{\mu\nu} \ = \ h_{MN}g^{MN} \ = \ 0\,, \qquad \nabla^\mu h_{\mu\nu} \ = \ 0\,.
\ee
The non-trivial components of the linearized Einstein equations take the form
\be
 \hat\nabla_\rho\hat\nabla^\rho h_{\mu\nu} + 2h_{\mu\nu} + \Delta_{{\rm S}_1^3}h_{\mu\nu} + \frac{c}{b}\cot^2\rho\nabla^2_{\rm CY_2}h_{\mu\nu} \ = \ 0\,,
\ee
where $\hat\nabla_\rho$ is the covariant derivative with respect to the AdS$_3$ metric and $\Delta_M$ is the scalar Laplace-Beltrami operator on the manifold $M$ (round 3-sphere), both with a unit radius. Written in terms of the explicit components of the metric from equation (\ref{eq:solsimp}) the above equations become
\be
\hat\nabla_\rho\hat\nabla^\rho h_{\mu\nu} + 2h_{\mu\nu} + \Delta_{S_1^3}{h}_{\mu\nu}  + \frac{c}{b}\frac{\cos^2\rho}{\sin^4\rho}\frac{\partial^2h_{\mu\nu}}{\partial \rho^2}+ \frac{c}{b}\frac{\cos3\rho - 3\cos \rho}{2\sin^5\rho}\frac{\partial h_{\mu\nu}}{\partial \rho} + \frac{c}{b}\frac{1}{\sin^2\rho}\Delta_{S_2^3}{h}_{\mu\nu} \ = \ 0\,.
\label{spin2eq}
\ee
Separation of variables
\be
h_{\mu\nu} \ = \ h_{{\mu\nu}}^{tt}(x^\sigma)H_{\ell_1,\ell_2}(\rho)Y_{{\rm S}_1^3}^{\ell_1}Y_{{\rm S}_2^3}^{\ell_2}\,,
\ee
where, as in~(\ref{harmexp}), we suppress the complete set of spin labels, results in the following equations for the compact part,
\be
\label{radeq}
\frac{c}{b}\frac{\cos^2\rho}{\sin^4\rho}\frac{d^2H_{\ell_1,\ell_2}}{d \rho^2}+ \frac{c}{b}\frac{\cos3\rho-3\cos \rho}{2\sin^5\rho}\frac{d H_{\ell_1,\ell_2}}{d \rho} \ = \  \left(-M^2 + {\ell_1(\ell_1+2)}+ \frac{c}{b}\frac{\ell_2(\ell_2+2)}{\sin^2\rho} \right)H_{\ell_1,\ell_2}\,.
\ee
In terms of the radial coordinate $r=\cos\rho$ this equation can be cast in the form
\be
r^2 H_{\ell_1,\ell_2}'' + 3r H_{\ell_1,\ell_2}'  -\left(\nu_\beta^2-1 + {\beta^2}r^2\right)H_{\ell_1,\ell_2} \ = \ 0\,,
\label{MBessel}
\ee
which can be reduced to the modified Bessel equation via a function redefinition. Here we introduced
\be
\nu_\beta^2 \ = \ (\ell_2+1)^2 - \beta^2\,, \qquad \beta^2 \ = \ \frac{b}{c}\left(M^2 - {\ell_1(\ell_1+2)}\right)\,.
\ee
The functions $p(r)$ and $q(r)$~(\ref{pandq}) defining the Sturm-Liouville problem~(\ref{canonicalSL}) in this case read
\be
p(r) \ = \ r^3\,,\qquad q(r) \ = \ r(1-r^2)\,.
\ee
The general solution of the above equation is in terms of the modified Bessel functions,
\be
H_{\ell_1,\ell_2} \ =\ C_1 \frac{K_{\nu_\beta}(\beta r)}{r} + C_2 \frac{I_{\nu_\beta}(\beta r)}{r}\,.
\ee
If $\nu_\beta$ is a real (we can assume it is positive), the function $K_{\nu_\beta}(r)\sim r^{-\nu_\beta}$, for small $r$, so the corresponding solution is neither normalisable, nor satisfying the admissible boundary conditions~(\ref{Neumann}) or~(\ref{Dirichlet}) at $r=0$. The second, linearly independent solution, has an admissible behaviour at $r\to 0$, however one cannot in general satisfy either boundary condition at $r=1$, because $I_{\nu_\beta}(\beta r)$ is a monotonous function for real $\beta$ and $\nu_\beta$. If $\beta$ is imaginary, then function $I_{\nu_\beta}(\beta r)$ is substituted by $J_{\nu_\beta}(\sqrt{-\beta^2}r)$, which is an oscillating function. Yet, the boundary conditions cannot be satisfied, because of the dependence of the parameter $\nu_\beta$ on $\beta$. This is a consequence of the positivity of the norm, as in the calculation~(\ref{NormPos}). The latter implies
\be
\label{Bound2}
\beta^2 \ \geq \ \ell_2(\ell_2+2)\frac{\int dr \ r|H|^2}{||H_{\ell_1,\ell_2}||^2}\ \geq \ 0\,.
\ee
Therefore there are no eigenvalues with imaginary $\beta$. Besides, the above relation implies that only states with $\ell_2=0$ can saturate the BPS bound of the ${\cal N}=(4,0)$ superconformal algebra, which is precisely $\beta^2\geq 0$ (see appendix~\ref{sec:smallN4}).

Hence, for real $\nu_\beta$ the only way to satisfy the boundary condition at $r=1$ is to set $\beta=\ell_2=0$, which makes $H_{\ell_1,\ell_2}$ a constant function. This case corresponds to BPS states with 
\be
\label{BPS1}
M^2 \ = \ \ell_1(\ell_1+2)\,, \qquad \Delta \ = \ \ell_1+2\,.
\ee
labelled by $\ell_1$ and containing the energy-momentum operator $\Delta=2$.

The remaining possibility to analyse, is the case of $\nu_\beta^2< 0$. This is expected as $\beta$ grows for fixed $\ell_2$ (larger $M^2$). At $\nu_\beta=0$ the Bessel functions change their behaviour to oscillating (as $r^{\nu_\beta}$ close to $r=0$) and, in fact, non-normalisable (with a logarithmically diverging norm). For this reason, we discard this continuum of values above
\be
\label{ContBound}
M^2 = \ell_1(\ell_1+2) +\frac{c}{b}(\ell_2+1)^2\,.
\ee
Altogether, on the background~(\ref{eq:solsimp}), the transverse traceless spin-two fluctuation $h_{\mu\nu}$ gives rise only to BPS branch~(\ref{BPS1}).


\subsection{Generalized AdS$_3\times$S$^3\times \mathbb{R}^4$ background}
\label{sec:spin2F}

Let us now move to the family of generalisations described by~(\ref{eq:CYsolgen}) with CY$_2$ fibred over S$^3_1$, as explained in section~\ref{sec:fibredsols}. We will focus on the  cases of the D5-O5 and O5-O5 geometries, introduced in sections~\ref{sec:D5O5withF} and~\ref{sec:O5O5}. The fibred geometry also falls into the class amenable to the general analysis of~\cite{Bachas:2011xa}, and the equation for the spin-two fluctuation of the metric~(\ref{spin2fluc}) can be cast in the form
\begin{multline}
\label{FibSpin2}
\frac{1}{L^2}\hat\nabla_\rho\hat\nabla^\rho h_{\mu\nu} + \frac{2}{L^2}h_{\mu\nu} + \frac{1}{L^2}\Delta_{S^3_1}h_{\mu\nu} + \frac{1}{\lambda^2h_5}\frac{\partial^2h_{\mu\nu}}{\partial r^2}+ \frac{3}{r}\frac{1}{\lambda^2h_5}\frac{\partial h_{\mu\nu}}{\partial r}  + \frac{1}{\lambda^2r^2 h_5}\Delta_{S^3_2}h_{\mu\nu} + 
\\ + \frac{4a_3r^2}{L^2h_5}\frac{\partial^2h_{\mu\nu}}{\partial \psi^2} - \frac{2}{\lambda^2h_5}{\cal A}^i \frac{\partial^2h_{\mu\nu}}{\partial \psi\partial x^i} \  = \ 0\,.
\end{multline}
As before, $\hat\nabla_\rho$ stands for the covariant derivative with respect to the metric of AdS$_3$ of unit radius and $\Delta$ are Laplacians on unit round 3-spheres. The contraction in the last term is calculated with respect to the unwarped CY$_2$ metric. Upon separation of variables, this equation reduces to equation~(\ref{spin2fibred}) for the factor depending on the compact space coordinates. As expected, the symmetry of the 3-sphere S$_1^3$ is broken down to $SU(2)\times U(1)$. This breaking is manifest in the second line of the above equation. 

Despite this generalisation, it is still convenient to use the basis of the scalar spherical harmonics on the round S$^3$ to expand the solution, that is we can use the substitution~(\ref{harmexp}). The last line of~(\ref{FibSpin2}) can be expressed as the Lie derivative with respect to connection ${\cal A}$, which is expanded as~(\ref{Aleftforms}) in terms of the left invariant forms on S$_2^3$, that is
\be
\nabla_{{\cal A}} Y^{\ell_2}_{\text{S}^3_2} \ \equiv \ {\cal L}_{\cal A}Y^{\ell_2}_{\text{S}^3_2} \ = \ \sum_a c_a {\cal L}_{K^R_a}Y^{\ell_2}_{\text{S}^3_2}\,, 
\ee
where $K^R_a$ are right invariant Killing vectors, dual to the left invariant forms on S$^3_2$, which act as the regular set of ladder operators (proper normalization of the harmonics is assumed),
\be
{\cal L}_{K^R_1\pm i K^R_2} Y^\ell_{m^R,m^L}  =  -i \sqrt{\frac{\ell}{4}(\ell+2)- m^R(m^R\pm 1)}\, Y^\ell_{m^R\pm1,m^L},\qquad {\cal L}_{K^R_3} Y^\ell_{m^R,m^L} =  i m^R Y^\ell_{m^R,m^L}.
\ee
Here we made explicit the quantum numbers of the spherical harmonics, with $(\ell^L,m^L)$ and $(\ell^R,m^R)$ being the pairs of $\ell$ and projection $-\ell/2\leq m\leq \ell/2$ quantum numbers with respect to SU(2)$_L\times$SU(2)$_R$ (for the scalar harmonics $\ell^L=\ell^R=\ell$). See appendix~\ref{sec:harmonics} for a summary of the properties of the scalar harmonics on a 3-sphere.

Substituting the expansion in the equation for the compact part and projecting onto harmonics one obtains
\begin{multline}
\label{FibEigenEq}
\frac{\partial^2H_{\ell_1,\ell_2}^{m^R_1,m^R_2}}{\partial r^2}+ \frac{3}{r}\frac{\partial H_{\ell_1,\ell_2}^{m^R_1,m^R_2}}{\partial r} - \frac{\ell_2(\ell_2+2)}{r^2} H_{\ell_1,\ell_2}^{m^R_1,m^R_2} - \frac{r^2}{2}\frac{\lambda^2{\cal C}}{L^2}(m_1^R)^2 H_{\ell_1,\ell_2}^{m^R_1,m^R_2} + 2c_3 m_1^R m_2^R H_{\ell_1,\ell_2}^{m^R_1,m^R_2} \\ -  2im_1^R\varpi_{\ell_2}^{m_2^R+ 1} H_{\ell_1,\ell_2}^{m^R_1,m^R_2+1} -   2im_1^R\varpi_{\ell_2}^{m_2^R- 1} H_{\ell_1,\ell_2}^{m^R_1,m^R_2-1}
\ = \  - \frac{\lambda^2h_5}{L^2}\left(M^2 - \ell_1(\ell_1+2)\right) H_{\ell_1,\ell_2}^{m^R_1,m^R_2}\,,
\end{multline}
where subscripts $1,2$ label the first and the second S$^3$, ${\cal C}$ is the parameter introduced in~(\ref{eq:PDElocalR4}), and we have to take into account that the action of the ladder operators in not diagonal. In particular, one obtains the off-diagonal coefficients
\be
\varpi_{\ell_2}^{m_2^R\pm 1} \ = \ -i \frac{c_1 \pm ic_2}{2}\sqrt{\frac{\ell_2}{4}(\ell_2+2)- m_2^R(m_2^R\pm 1)}\,.
\ee
First of all, we are interested in the zero modes of equation~(\ref{FibRadEq}), $H_{\ell_1,\ell_2}^{m^R_1,m^R_2}=\text{const}$. In particular the stress-energy tensor mode $\ell_1=\ell_2=M=0$ remains on this branch.

Let us consider the case $\ell_2=0$. Then the off-diagonal terms drop out and the equation can be cast in the form
\be
\label{FibRadEq}
\frac{\partial^2H_{\ell_1}^{m^R_1}}{\partial r^2}+ \frac{3}{r}\frac{\partial H_{\ell_1}^{m^R_1}}{\partial r} + \left(\alpha_1+\frac{\alpha_2}{r^2}-\alpha_3 r^2\right)H_{\ell_1}^{m^R_1} \ = \ 0\,,
\ee
with
\be
\alpha_1 \ = \ a_1\beta^2, \qquad 
\alpha_2 \ = \ a_2\beta^2 \,, \qquad \alpha_3 \ = \ a_3\beta^2 + 4a_3\frac{\lambda^2}{L^2}(m_1^R)^2\,.
\ee
In the above formulae 
\be
\beta^2\ =\ \frac{\lambda^2}{L^2}\left(M^2-\ell_1(\ell_1+2)\right)
\ee
and $a_1$, $a_2$, $a_3={\cal C}/8>0$ are the parameters of the warp factor~(\ref{fibredwarp}). From~(\ref{pandq}) we find
\be
\label{pandqfib}
p(r) \ = \ r^3 \,,\qquad q(r)\ = \ r^3 h_5 \ = \ r^3\left(a_1+\frac{a_2}{r^2}-a_3 r^2\right).
\ee
With this we can derive a bound on the eigenvalues,
\be
\label{Bound3}
\beta^2 \ \geq \ 4a_3\frac{\lambda^2}{L^2}(m_1^R)^2\frac{\int dr \ r^5|H|^2}{||H_{\ell_1,\ell_2}||^2}\ \geq \ 0\,.
\ee
The bound is only compatible with the BPS bound of ${\cal N}=(4,0)$ superconformal algebra if $m_1^R=0$. While in the analysis of section~\ref{sec:spin2noF} we observed a degeneracy of the BPS spectrum with respect to the projection of momentum $\ell_1$, due to the full SO(4) symmetry, here only the states with zero projection are BPS. In particular, this excludes the states with odd $\ell_1$. Another feature of the generalized background is that non-BPS states exist for some values of the parameters, as we will see.

We can write the general solution to equation~(\ref{FibRadEq}),
\be
\label{D5O5GenSol}
H_{\ell_1}^{m_1^R} \ = \ r^{\alpha-1}e^{-\frac{\sqrt{\alpha_3}}{2} r^2}\left\{C_1 L_\nu^\alpha(\sqrt{\alpha_3}r^2)+C_2U(-\nu,1+\alpha,\sqrt{\alpha_3}r^2)\right\}\,,
\ee
where $L_\nu^a(x)$ are the generalised Laguerre polynomials, $U$ is the hypergeometric $U(a,b,x)$ function (Kummer's function of the second kind) and the parameters are 
\be
\label{HeqParams}
\alpha \ = \sqrt{1-\alpha_2}\,, \qquad \nu \ = \  \frac{\alpha_1\sqrt{\alpha_3}-2\alpha_3(1+\alpha)}{4\alpha_3}\,.
\ee
For $r\to 0$ the asymptotic form of the solution is
\be
H_{\ell_1}^{m_1^R} \ \sim \ r^{-1\pm\sqrt{1-\alpha_2}}\,.
\ee
Note that the sign of $\alpha_2$ distinguishes the D5-O5~(\ref{D5O5warp}) and O5-O5~(\ref{eq:O5O5}) configurations. For $\alpha_2>0$ we are in the D5-O5 case, of which the analysis of section~\ref{sec:spin2noF} has treated a particular representative. As in section~\ref{sec:spin2noF}, we observe a cut, where the solutions become infinitely oscillating and unnormalisable. In the generalised background this happens when
\be
\label{FibThresh}
M^2\ = \ \ell_1(\ell_1+2) + \frac{8L^2}{b_1^2b_2^2\lambda^2{\cal C}} \ = \ \ell_1(\ell_1+2) + \frac{N_5}{n}\,,
\ee
where we used parametrisation~(\ref{h5D5O5O5}) of the warp factor. Now, in the general D5-O5 case a finite number of eigenvalues is possible below the threshold. 

\begin{figure}[t]
\centering
    \includegraphics[width=0.45\linewidth]{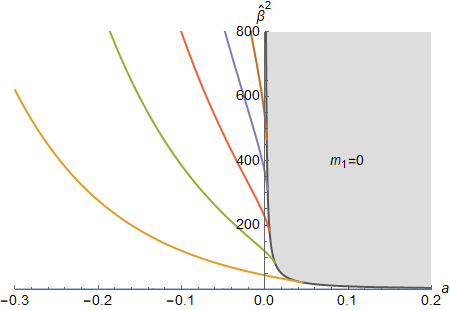}
    \hfill
    \includegraphics[width=0.45\linewidth]{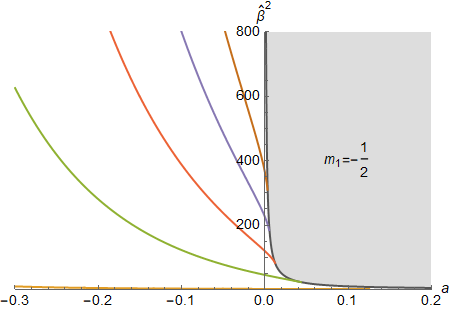}\\
    \includegraphics[width=0.45\linewidth]{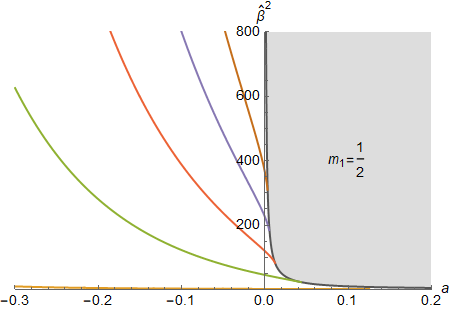}
    \hfill
    \includegraphics[width=0.45\linewidth]{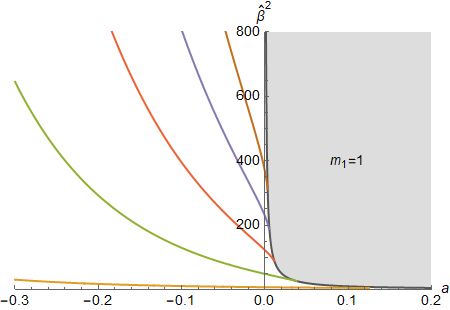}
    \caption{Spectrum of the first five eigenvalues $\hat{\beta}^2$ of equation~(\ref{DimLessSpin2}) for $m_1^R=0,-1/2,1/2,1$ (as labelled). The case $m_1^R=0$ contains an additional line at $\hat{\beta}^2=0$ representing the BPS families of spin-two states in the backgrounds of section~\ref{sec:fibredsols} labelled by $\ell_1$. The remaining curves represent non-BPS families with $\ell_2=0$ and any $\ell_1\geq 2m_1^R$. The parameter $a$ interpolates between the O5-O5 configuration ($a=-\hat{b}_1^2\leq 0$) and the D5-O5 one ($a=\hat{b}_1^2>0$). The O5-O5 configuration shows an unbounded discrete spectrum. For the D5-O5 case the discrete spectrum is bounded from above by a cut separating the continuum of unnormalizable solutions (shaded region).}
    \label{fig:eigsl20}
\end{figure}

The spectrum of the eigenvalues is defined by zeroes of the above special functions and of their derivatives. The latter can be determined numerically. It is convenient to make a rescaling of the parameters,
\be
\label{rrescaling}
r^2\ = \ \frac{L}{\sqrt{a_3}\lambda}\hat{r}^2\,, \qquad b_1^2\ = \ \frac{L}{\sqrt{a_3}\lambda}\hat{b}_1^2\,, \qquad b_2^2\ = \ \frac{L}{\sqrt{a_3}\lambda}\hat{b}_2^2\,, \qquad \beta^2=\frac{\lambda^2}{L^2}\hat{\beta}^2\,.
\ee
We can also set $\hat{b}_2=1$ without loss of generality. Equation~(\ref{FibRadEq}) then takes the following form
\be
\label{DimLessSpin2}
\frac{\partial^2H_{\ell_1}^{m^R_1}}{\partial \hat{r}^2}+ \frac{3}{\hat{r}}\frac{\partial H_{\ell_1}^{m^R_1}}{\partial \hat{r}}+\hat{\beta}^2\frac{(\hat{b}_1^2+\hat{r}^2)(1-\hat{r}^2)}{\hat{r}^2} H_{\ell_1}^{m^R_1} - 4 \hat{r}^2(m_1^R)^2 H_{\ell_1}^{m^R_1} \ = \ 0\,,
\ee
The normalisable solution to this equation is the one in~(\ref{D5O5GenSol}), in terms of the generalised Laguerre polynomial. 
\be
\label{FibSol1}
H_{\ell_1}^{m^R_1} \ = \ \hat{r}^{\alpha-1}e^{-\frac{\varkappa}{2} \hat{r}^2} L_\nu^\alpha(\varkappa\hat{r}^2)\,,
\ee
where now
\be
\label{YeqParams}
\varkappa \ = \ \sqrt{\hat\beta^2+4(m_1^R)^2}\,, \qquad \alpha \ = \sqrt{1-\hat{b}_1^2\hat\beta^2}\,, \qquad \nu \ = \  \frac{\hat\beta^2(1-\hat{b}_1^2)}{4\varkappa}-\frac{1+\alpha}{2}\,.
\ee
As was noticed in the previous analysis, the BPS spectrum is only compatible with the Neumann boundary conditions at the position of O5, that is at $r=b_2$ ($\hat{r}=1$). Therefore we impose the condition
\be
\partial_{\hat{r}} H_{\ell_1}^{m^R_1}\Big|_{\hat{r}=1} \ = \ 0\,
\ee
on solution~(\ref{FibSol1}). The spectrum of the eigenvalues is shown in figure~\ref{fig:eigsl20} as a function of the parameter $a$, which coincides with $\hat{b}_1^2$ for $a> 0$ ($a$ is the same as the rescaled parameter $a_2$ in the warp factor). As anticipated, there is a finite number of eigenvalues below a cut with the continuum of unnormalizable solutions. The profile of the cut is defined by the bound~(\ref{FibThresh}). For example, for $m_1^R=0$ and $a>0.04555$ one finds no non-BPS states below the continuum threshold. Conversely the number of eigenvalues increases as $a$ approaches zero.

The exact limit $\hat{b}_1=a=a_2=0$ is in fact the intermediate case with the solution bounded between a single O5 plan and a regular zero, while for $a<0$ one is in the O5-O5 configuration (section~\ref{sec:O5O5}), so that $a=-\hat{b}_1^2$. In the latter case equation~(\ref{DimLessSpin2}) is modified as follows,
\be
\frac{\partial^2H_{\ell_1}^{m^R_1}}{\partial \hat{r}^2}+ \frac{3}{\hat{r}}\frac{\partial H_{\ell_1}^{m^R_1}}{\partial \hat{r}}+\hat{\beta}^2\frac{(\hat{b}_1^2-\hat{r}^2)(\hat{r}^2-1)}{\hat{r}^2} H_{\ell_1}^{m^R_1} - 4 \hat{r}^2(m_1^R)^2 H_{\ell_1}^{m^R_1} \ = \ 0\,,
\ee
and its regular solution reads
\be
H_{\ell_1}^{m_1^R} \ = \ \hat{r}^{\bar{\alpha}-1}e^{\frac{\varkappa}{2} \hat{r}^2} L_{\bar{\nu}}^{\bar{\alpha}}(-\varkappa\hat{r}^2)\,,
\ee
with modified parameters
\be
\bar{\alpha} \ = \sqrt{1+\hat{b}_1^2\hat\beta^2}\,, \qquad \bar{\nu} \ = \ - \frac{\hat\beta^2(\hat{b}_1^2+1)}{4\varkappa} - \frac{(1+\bar{\alpha})}{2}\,.
\ee
Again, we stick to Neumann boundary conditions, and the results of the analysis can be found in figure~\ref{fig:eigsl20} as a continuation of the spectrum to negative $a$. In this branch the spectrum is discrete and unbounded.

Now let us turn on $\ell_2$. For $\ell_2\neq 0$ we will restrict our analysis to $c_1=c_2=0$, with non-vanishing $c_3$. This keeps eigenmode equations~(\ref{FibEigenEq}) diagonal. The equations also take the form~(\ref{FibRadEq}), but this time with
\be
\label{alphap2}
\alpha_1 \ = \ a_1\beta^2 +2c_3m_1^Rm_2^R\,, \qquad 
\alpha_2 \ = \ a_2\beta^2 - \ell_2(\ell_2+2) \,, \qquad \alpha_3 \ = \ a_3\beta^2 + \frac{c_3^2}{4}(m_1^R)^2\,,
\ee
where we used the relation between $a_3$ and $c_3$ provided by equations~(\ref{eq:PDElocalR4}) and~(\ref{fibredwarp}). Before we discuss the spectrum let us check the bound on the eigenvalue $\beta^2$. The same type of argument as before yields
\be
\beta^2 \ \geq \ \frac{1}{||H_{\ell_1,\ell_2}||^2}\int dr \ r^3|H|^2\left(\frac{\ell_2(\ell_2+2)-4(m_2^R)^2}{r^2}+\left(\frac{c_3m_1^Rr}{2}-\frac{2m_2^R}{r}\right)^2\right),
\ee
which generalises the previous bounds~(\ref{Bound2}) and~(\ref{Bound3}) and is compatible with the BPS bound of the superalgebra $\beta^2\geq 0$. The BPS bound can only be saturated by configurations with $\ell_2=0$ and $m_1^R=0$.

\begin{figure}[ht]
    \centering
    \includegraphics[width=0.45\linewidth]{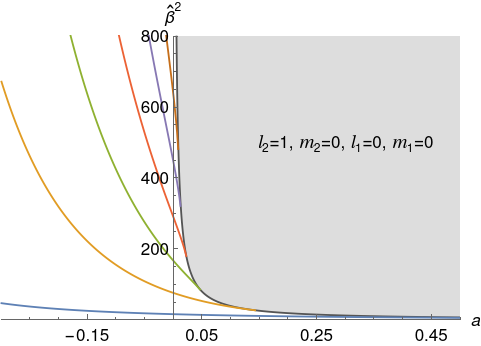}
    \hfill
    \includegraphics[width=0.45\linewidth]{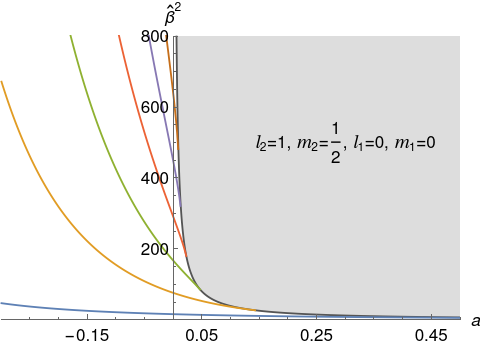}\\
    \includegraphics[width=0.45\linewidth]{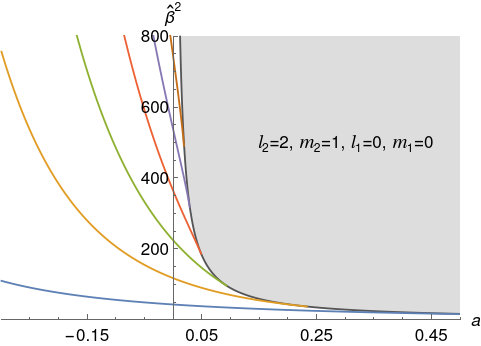}
		\hfill
		\includegraphics[width=0.45\linewidth]{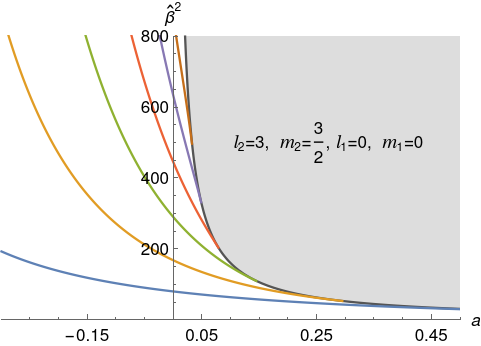}
    \caption{Spectrum of the first six non-BPS families of eigenvalues $\hat{\beta}^2$ of equation~(\ref{FibRadEq}) with parameters given by~(\ref{alphap2}), $\ell_1=0$ and a few choices for $\ell_2$ and $m_2^R$ (as labelled). Parameter $a$ is defined as in figure~\ref{fig:eigsl20}.}
    \label{fig:betab1plot2}
\end{figure}

\begin{figure}[ht]
    \centering
    \includegraphics[width=0.45\linewidth]{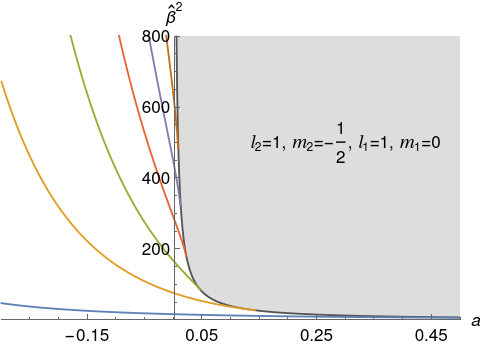}
    \hfill
    \includegraphics[width=0.45\linewidth]{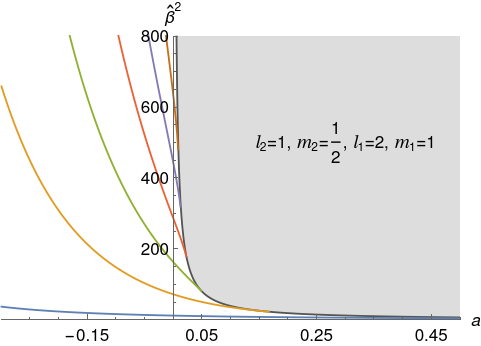}\\
    \includegraphics[width=0.45\linewidth]{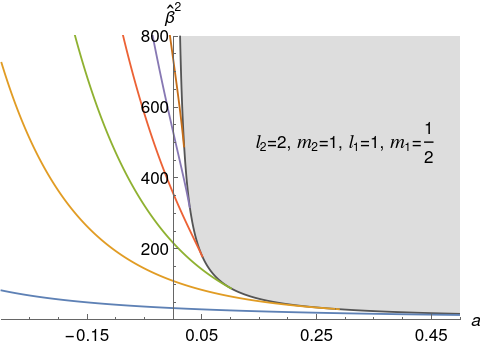}
    \hfill
    \includegraphics[width=0.45\linewidth]{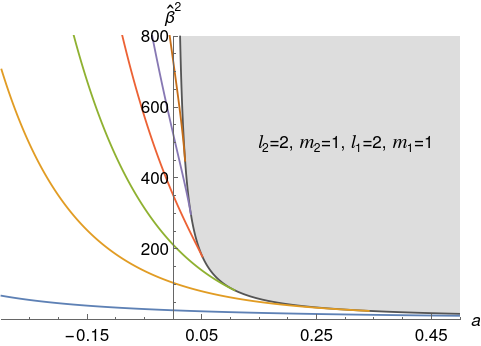}
    \caption{Spectrum of the first six non-BPS families of eigenvalues $\hat{\beta}^2$ of equation~(\ref{FibRadEq}) with parameters given by~(\ref{alphap2}) and a few choices of $\ell_1\neq 0$, $m_1^R$, $\ell_2$ and $m_2^R$ (as labelled). Parameter $a$ is defined as in figure~\ref{fig:eigsl20}.}
    \label{fig:betab1plot3}
\end{figure}

The asymptotic solution at $r\to 0$ has the form
\be
 H_{\ell_1,\ell_2}^{m^R_1,m^R_2} \sim r^{-1\pm \sqrt{(\ell_2+1)^2-a_2\beta^2}}\,.
\ee
The branch with minus sign is not normalisable with $q$ from equation~(\ref{pandq}). The branch with plus sign is only normalisable if $a_2\beta^2<(\ell_2+1)^2$. For $a_2>0$, which is the case of the D5-O5 configuration, the normalisable spectrum is bounded,
\be
0\leq \beta^2 < \frac{(\ell_2+1)^2}{a_2} \ = \ \frac{N_5}{n}(\ell_2+1)^2\,,
\ee
which generalises the bound~(\ref{FibThresh}) to arbitrary $\ell_2$. For $a_2<0$ (O5-O5) the upper bound disappears and the spectrum is only bounded from below. The general solution is the same as equation~(\ref{D5O5GenSol}) but with the updated values of the parameters~(\ref{alphap2}). The generalised Laguerre polynomial represents the normalisable branch. The spectrum for some choices of quantum numbers with $\ell_2\neq 0$ is shown in figures~\ref{fig:betab1plot2} and~\ref{fig:betab1plot3}.

\section{SU(2)$_R$ current mode}
\label{sec:spin1}

The stress-energy tensor is the highest component of the short superconformal multiplet (appendix~\ref{sec:smallN4}), which also contains the SU(2) R current triplet as the remaining (lowest) bosonic components. In this section we will construct and analyse the dual supergravity mode. We shall only consider the simplest background~(\ref{eq:solsimp}) of the family~(\ref{eq:CYsolgen}).

The supergravity mode dual to the SU(2) R currents should correspond to the fluctuation of the metric that is a vector on each of AdS$_3$ and S$_1^3$.
\be
\label{vectorR}
\delta g \ = \ 2R_{\mu a}dx^\mu d\phi^a \ = \ 2 A_\mu(x^\mu,y^i) Y_{a,1}^{(\ell_1, \pm 1)} dx^\mu d\phi^a\,,
\ee
where $x^\mu$ and $y^i$ are AdS$_3$ and CY$_2$ coordinates and $\phi^a$ and $Y_{a,1}^{(\ell_1,\pm1)}$ are the coordinates and vector harmonics on S$_1^3$ respectively\footnote{In this section we use Greek indices $\mu$, $\nu$, $\rho$, \ldots for the AdS$_3$ directions, Latin indices $a$, $b$, $c$, \ldots for the directions along S$^3$ and $i$, $j$, $k$, \ldots for the CY$_2$.}. The latter are labeled by the sum $\ell_1=j_1+\bar{j}_1$ of the two spins of SU(2)$\times$SU(2)$\simeq$SO(4), while the difference is fixed to $\pm 1$.

It turns out that perturbing the above metric component alone does not lead to a consistent set of linearised equations. The minimal consistent ansatz requires  one to perturb the Ramond 3-form as well,
\be
\label{vectorC}
\delta C_2 \ = \ C_{\mu a}\, dx^\mu\wedge dx^a\,.
\ee
The perturbations $R_{\mu a}$ and $C_{\mu a}$ do provide a well-defined system of linearised equations.\footnote{We label the modes following conventions of~\cite{Eberhardt:2017fsi,Deger:1998nm}.}

Note that the left and right vector spherical harmonics on S$^3$ of unit radius satisfy
\be
\label{SqrtHarm}
\hat{\epsilon}_a^{~bc}\partial_b Y_c^{(\ell,\pm 1)} \ = \ \pm (\ell+1)Y_{a}^{(\ell,\pm 1)} \,,
\ee
where the indices are raised with the S$^3$ metric and $\hat{\epsilon}$ is the covariant Levi-Civita tensor. This equation reflects the fact that the harmonics form a representation of SU(2). It can also be viewed as a ``square root" of the Laplacian eigenvalue equation. Indeed, applying the differential operator twice  on the left hand side yields
\be
\label{S3VLap}
\hat\nabla^d\hat\nabla_dY_a^{(\ell,\pm 1)} \ = \ (2-(\ell+1)^2)Y_a^{(\ell,\pm 1)}\,,
\ee
which is the correct action of the Laplacian on the harmonics. As before we dress with hats the operators defined with respect to constant curvature unit radius metric on either S$^3$ or AdS$_3$. 

Similar relations are satisfied by the AdS$_3$ wavefunctions. The appropriate basis is that of the representations of SL(2), satisfying an analog of~(\ref{SqrtHarm}),
\be
\label{SqrtAdS3}
\hat\epsilon_\mu^{~\rho\eta}\partial_\rho R_{\eta a}^{(\Delta,\pm 1)} \ = \ \mp {(\Delta-1)} R_{\mu a}^{(\Delta,\pm 1)} \,.
\ee
Different signs correspond to the choice of the left ($h-\bar h=1$) or the right ($h-\bar h=-1$) mover sectors, $\Delta=h+\bar h$. Then 
\be
\label{AdSVLap}
\hat\nabla^\rho\hat\nabla_\rho R_{\mu a}^{(\Delta,\pm 1)} - \hat\nabla^\rho\hat\nabla_\mu R_{\rho a}^{(\Delta,\pm 1)} \ = \ {(\Delta-1)^2} R_{\mu a}^{(\Delta,\pm 1)}\,.
\ee
The latter is the equation for the vector field of mass $M^2=(\Delta-1)^2$ in AdS$_3$ measured in the units of the inverse radius, which is the correct relation between the AdS mass and the conformal dimension $\Delta$ of the operator dual to a vector field.

Imposing the additional transversality condition $\hat\nabla^\rho R_{\rho a}=0$ casts (\ref{AdSVLap}) in the form
\be
\hat\nabla^\rho\hat\nabla_\rho R_{\mu a}^{(\Delta,\pm 1)}  \ = \ \left((\Delta-1)^2-2\right) R_{\mu a}^{(\Delta,\pm 1)}\,,
\ee
which is the analog of~(\ref{S3VLap}).

One can check that linearized fluctuations~(\ref{vectorR}) and~(\ref{vectorC}) only contribute to Einstein and $F_3$ equations of motion~\cite{inprep}, which in string frame yield
\begin{multline}
\label{LinVec1}
- \nabla_P\nabla^P R_{MN} + \nabla^P \nabla_MR_{NP} +
\nabla^P\nabla_N R_{MP}  + 2\left(\nabla_PR_{MN}-\nabla_N R_{PM}-\nabla_MR_{PN}\right)\partial^P\Phi  \\ + \frac{e^{2\Phi}}{12}F_{PQR}F^{PQR}R_{MN} - \frac{e^{2\Phi}}{2}S_{MPQ}F_N^{~PQ} - \frac{e^{2\Phi}}{2}S_{NPQ}F_M^{~PQ} \ = \ 0\,.
\end{multline}
and
\be
\label{LinVec2}
\partial_M\left(\sqrt{g}S^{MRS}\right) 
- \partial_M\left(\sqrt{g} R^{MP}F_P^{~RS}\right) 
- \partial_M\left(R^{R}_{~P}\right)\sqrt{g}F^{PSM} + \partial_M\left(R^{S}_{~P}\right)\sqrt{g}F^{PRM}   =  0\,. 
\ee 
Here $S_{MNP}$ are the components of the 3-form field strength tensor of the 2-form $C_{\mu a}$.

In fact, $R_{\mu a}$ also contributes to the dilaton equation of motion, but this contribution
\be
\hat\nabla^\mu\hat\nabla_aR_{\mu}^{~a} \ = \ 0\,,
\ee
only affects the gauge dependent part of the equations. If we fix the gauge with the local transversality conditions
\be
\hat\nabla^\nu R_{\nu a} \ = \ 0\,, \qquad \hat\nabla^b R_{\mu b} \ = \ 0\,, \qquad \hat\nabla^\nu C_{\nu a} \ = \ 0\,, \qquad \hat\nabla^b C_{\mu b}\ = \ 0\,, 
\ee
then the dilaton and all the components of Einstein and $F_3$ equations decouple, with the exception of the components $\mu a$, that is those with one leg along AdS$_3$ and the other along the first S$^3$. In a separate work~\cite{inprep} we show that the considered fluctuations also decouple from space-time filling sources, ie of the type contained in the solutions we consider.

The equations become simpler if the metric perturbation is of the form $R_{MN} = e^{\Phi}V_{MN}$. In terms of $V_{\mu a}$ we get
 \begin{multline}
 \frac{1}{L^2}\hat\nabla^\rho \hat\nabla_\rho V_{\mu a} - \frac{1}{L^2}\hat\nabla^\rho\hat\nabla_\mu V_{\rho a}  
+ \frac{1}{L^2}\hat\nabla^c \hat\nabla_c V_{\mu a} - \frac{1}{L^2}\hat\nabla^c\hat\nabla_a V_{\mu c}   + \frac{1}{h_5\lambda^2}\nabla_{\text{CY}_2}^2 V_{\mu a} -
\\ - \frac{2}{L^2}\left( \hat\epsilon_{\mu\nu\lambda} \partial^\lambda C^\nu_{~a} - \hat\epsilon_{abc} \partial^c C_\mu^{~b}\right)\ = \ 0\,,
\label{Veq}
\end{multline}
where again $\hat\nabla$ and $\hat\epsilon$ with Greek or Latin indices refer to the covariant derivatives and covariant Levi-Civita symbols on AdS$_3$ and S$^3$ of unit radii respectively, and
\begin{multline}
  \frac{1}{L^2}\hat\nabla^\rho \hat\nabla_\rho C_{\mu a} - \frac{1}{L^2}\hat\nabla^\rho\hat\nabla_\mu C_{\rho a}  
+ \frac{1}{L^2}\hat\nabla^c \hat\nabla_c C_{\mu a} - \frac{1}{L^2}\hat\nabla^c\hat\nabla_a C_{\mu c}  + \frac{1}{h_5\lambda^2}\nabla_{\text{CY}_2}^2 C_{\mu a} +
\\  - \frac{2}{L^2}\left(\hat\epsilon_{\mu\nu\lambda} \partial^\lambda V^\nu_{~a} - \hat\epsilon_{abc} \partial^c V_\mu^{~b}\right)\ = \ 0\,.
\label{Ceq}
\end{multline}

To diagonalise the coupled system~(\ref{Veq})-(\ref{Ceq}) we first pass to the modes $V^{\pm}_{\mu a}=V_{\mu a}\pm C_{\mu a}$.
 \begin{multline}
 \frac{1}{L^2}\hat\nabla^\rho \hat\nabla_\rho V_{\mu a}^\pm - \frac{1}{L^2}\hat\nabla^\rho\hat\nabla_\mu V_{\rho a}^\pm + \frac{1}{L^2}\hat\nabla^c \hat\nabla_c V_{\mu a}^\pm - \frac{1}{L^2}\hat\nabla^c\hat\nabla_a V_{\mu c}^\pm   + \frac{1}{h_5\lambda^2}\nabla_{\text{CY}_2}^2 V_{\mu a}^\pm \mp 
\\ \mp \frac{2}{L^2}\left( \hat\epsilon_{\mu\nu\lambda} \partial^\lambda {V^{\pm\nu}}_{a} - \hat\epsilon_{abc} \partial^c V_\mu^{\pm b}\right)\ = \ 0\,.
\end{multline}
There is a complication in the above equations since the last terms couple different AdS$_3$ components and different S$^3$ components. However, the harmonic expansion brings the equations to the diagonal form through relations~(\ref{SqrtHarm}) and~(\ref{SqrtAdS3}). Expanding
\be
V^\pm_{\mu a} \ = \ \sum\limits_{\Delta,\ell}R_{\mu}^{(\Delta,\pm1)}(x^\mu)Y_a^{(\ell,\pm 1)}(y^a)V_{\Delta,\ell}^{\pm(\pm,\pm)}(y^i)\,,
\ee
one obtains a set of equations for the unknown functions of the CY$_2$ coordinates $y^i$,
 \begin{multline}
 \frac{1}{h_5\lambda^2}\nabla_{\text{CY}_2}^2 V_{\Delta,\ell_1}^{\pm(\pm,\pm)} + \frac{(\Delta-1)^2}{L^2}V_{\Delta,\ell_1}^{\pm(\pm,\pm)} -  \frac{(\ell_1+1)^2}{L^2}V_{\Delta,\ell_1}^{\pm(\pm,\pm)} \\ \mp \frac{2}{L^2}\left[\mp (\Delta-1) \mp (\ell_1+1)\right]V_{\Delta,\ell_1}^{\pm(\pm,\pm)}\ = \ 0\,,
 \label{VpmEq}
\end{multline}
which determines the spectrum. Here the choice of the signs in the last term of the equation follows the order of the choice in the function $V^{\pm(\pm,\pm)}_{\Delta,\ell_1}$. The first choice of the sign is the choice of the mode $V^{\pm}_{\mu a}$ of the vector perturbation of the metric and $F_3$. The other two correspond to the choice of the left or the right mover sectors in SO(2,2) and SO(4) respectively. 

Equation~(\ref{VpmEq}) is of the same form as equation~(\ref{eq:spin2eigenfunctioneq}) derived for the spin two mode if we set
\be
M^2L^2 \ = \ \Delta(\Delta-2) -1 \mp 2\left[\mp (\Delta-1) \mp (\ell_1+1)\right]\,,
\ee
so all that is said about the analysis of the spectrum of the spin two mode applies to the vector modes, modulo a constant shift, at least in the case of the background solution~(\ref{eq:solsimp}). It is then sufficient to restrict our discussion to the case of zero modes of CY$_2$, which correspond to BPS states.

For the R symmetry current the simple relation with the spin two mode is expected, because both modes enter the same short superconformal multiplet, so the linearized equations too should be related by a simple supersymmetric transformation~\cite{Dymarsky:2007zs,Dymarsky:2008wd}. The R symmetry is associated with one of the SU(2) factors of SO(4), say the one labeled by $j$ (as opposed to $\bar{j}$). Hence we have to choose one of the signs of $(\ell_1+1)$ in the above equation to correspond to the SU(2) R current. Choosing $V^+$ mode with $j-\bar{j}=1$ (choosing the first sign for $(\ell_1+1)$) one obtains the zero mode relation
\be
{(\Delta-1)^2} - (\ell_1+1)^2 \pm  2(\Delta-1) + 2(\ell+1)\ = \ 0\,.
\ee
The first choice of the sign corresponds to $\Delta=\ell_1$ (we ignore the non-unitary solution $\Delta= -\ell_1$). For the second choice of the sign, the solution is $\Delta = \ell_1+2$ (ignore $\Delta=-\ell_1+2$).

If we choose $\bar{j}-j=1$ then
 \be
\Delta(\Delta-2) \pm 2(\Delta-1) \ = \ \ell_1(\ell_1+4)+2\,.
\ee
One finds $\Delta = \ell_1+2$ for the first choice of the sign and $\Delta \ = \ \ell_1 + 4$ for the second. Again, we omit the non-unitary branches.

The analysis for the $V^-$ mode duplicates the above spectrum, but exchanges the left-right sector assignment. The result is summarised in table~\ref{tab:Rvectormodes}. The pair of states with $\Delta=\ell_1$ should correspond to the left and right components of the conserved SU(2) R current ($\ell_1=1$), which are the lowest components of the short spin two multiplet, and an associated tower of operators labeled by $\ell_1$. There are also two pairs of $\Delta=\ell_1+2$ towers that can be either  highest components of some short multiplets or intermediate components of long multiplets. There is also a pair of heavy $\Delta=\ell_1+4$ towers, which must belong to a long multiplet. In table~\ref{tab:Rvectormodes} we label the states by the mode index $V^{\pm}$, by the difference of conformal dimensions $h-\bar h=\pm 1$ ($\Delta=h+\bar h$) and by the difference of S$^3$ spins $j-\bar j=\pm 1$ ($\ell_1=j+\bar j$).

\begin{table}[h]
    \centering
    \begin{tabular}{||c|c|c|c|c||c|c|c|c|c||}
    \hline \Trule\Brule Operator & Mode & $h-\bar h$ & $j-\bar j$ & $\Delta$& Operator & Mode & $h-\bar h$ & $j-\bar j$ & $\Delta$ \\
        \hline \Trule\Brule R current, $J_\mu^a$ & $V^+$ & +1 & +1 & 1 & a/hol. $\bar{J}_\mu^a$ & $V^-$ & -1 & -1 & 1 \\
         \hline \hline \Trule\Brule $O_\mu^R$ & $V^+$ & +1 & +1 & $\ell_1$ &  $\bar{O}_\mu^L$ & $V^-$ & -1 & -1 & $\ell_1$ \\
         \Trule\Brule $O_\mu^{(1)}$ & $V^-$ & +1 & -1 & $\ell_1+2$ &  $\bar{O}_\mu^{(1)}$ & $V^+$ & -1 & +1 & $\ell_1+2$ \\
        \Trule\Brule $O_\mu^{(2)}$ & $V^+$ & +1 & -1 & $\ell_1+2$ & $\bar{O}_\mu^{(2)}$ & $V^-$ & -1 & +1 & $\ell_1+2$\\
         \Trule\Brule $O_\mu^{(3)}$ & $V^-$ & +1 & +1 & $\ell_1+4$ & $\bar{O}_\mu^{(3)}$ & $V^-$ & -1 & -1 & $\ell_1+4$\\
         \hline
    \end{tabular}
    \caption{Operators described by the vector perturbations $R_{\mu a}$ and $C_{\mu a}$. The first line corresponds to the special cases of the SU(2)$_{\rm R}$ current. Here $\Delta=h+\bar h$, $\ell_1=j+\bar j$. }
    \label{tab:Rvectormodes}
\end{table}


\section{Comments on string theory for new ${\cal N}=(4,0)$ AdS$_3$ solutions} 
\label{sec:stringsAdS3}

It is natural to ask if the supergravity solutions discussed in the present work have string theory completions, and whether one can learn something new about them from a worldsheet construction. In this section, we make some comments regarding the question of which string theory would lead to the new ${\cal N}=(4,0)$ AdS$_3$ solutions presented in this paper.\\

Well-known examples of superstrings on AdS$_3$ are the type IIB on AdS$_3\times$S$^3\times$M$_4$, where M$_4$ is either $\mathbb{T}^4$ or K$3$. 
These backgrounds are obtained by considering the near horizon limit of some branes.  The new ingredients appearing in the supergravity solutions of this paper are 
additional extended objects: D-branes and O-planes (or their S-dual relatives, NS fivebranes and ONS5-planes\footnote{In the rest of the paper, supergravity solutions with only RR 3-form flux are considered. It is of course simple to map these to pure NS solutions via S-duality. Nevertheless, the study of NS fivebranes differ from that of D-branes since only the latter can be analyzed perturbatively.}). Besides, the solutions have ${\cal N}=(4,0)$ symmetry, and the reduction from ${\cal N}=(4,4)$ to ${\cal N}=(4,0)$ is a consequence of adding
these extra sources. 

One can start considering the presence of O5-planes extended in the AdS$_3\times $S$^3$ directions. In type IIB string theory, these objects appear as the fixed loci of elements of a group ${\mathcal{G}}=G_1\times \Omega_w G_2$ acting on both the worldsheet and the target space \cite{HoravaOrbifold}. Here, $\Omega_w$ is the parity  transformation of the worldsheet which exchanges left- and right-moving sectors of closed strings; for instance, in type IIB superstring in a flat background, the graviton $G_{MN}$, dilaton $\phi$, and R-R two-form $C_{MN}$ are all even under $\Omega_w$, while the NS-NS two-form $B_{MN}$, the axion $C$ (R-R scalar), and R-R four-form with self-dual field strength $C_{MNPQ}$, are odd under $\Omega_w$. On the other hand, the groups $G_1$ and $G_2$ only act on the target space. These can be any discrete isometry of the space, where in particular the unique non-trivial element of $G_2$ is taken to be an involution $\mathbb{Z}_2$ denoted by $\sigma$. The orientifold is constructed by suitably projecting out states in the type IIB spectrum that are non-invariant under ${\mathcal{G}}$, while adding, if necessary, extra states belonging to {\it twisted sectors} which are states that satisfy periodic boundary conditions up to conjugation of an element of $G_1$.

The fixed loci of the set of elements of ${\mathcal{G}}$ acting on the target space correspond to the position of extended objects. For example, if ${\mathcal{G}}$ contains only the element $\Omega_w$, its action is trivial on the target space, thus its fixed locus is the entire target space: it will give rise to spacetime-filling O9-planes. These objects do not appear in the new supergravity solutions considered here; this means that the orientifold group should be of the form $G_1\times \{\mathbbm{1}, \Omega_w \sigma\}$.

Taking AdS$_3\times $S$^3\times $M$_4$ as target space, the aforementioned O5-planes arise when $\sigma$ is an involution of M$_4$, whose fixed locus is zero-dimensional. Locally, one can take
\begin{equation}
	\sigma: (y^6,y^7,y^8,y^9) \ \mapsto \ (-y^6,-y^7,-y^8,-y^9) \, . 
\end{equation}
Thus if M$_4=\mathbb{T}^4$, then there are sixteen fixed points, that is, sixteen O5-planes extended in the AdS$_3\times$S$^3$ directions and located at points $(a^6, ..., a^9)$ where $a^i =0$, or $a^i=\pi R$ ($R$ being the radius of $\mathbb{T}^4$). In the limit $R\rightarrow\infty$, the torus is decompactified to $\mathbb{R}^4$ with the origin as the unique fixed point. For K$3$ as internal manifold, $\sigma$ can be taken to be a holomorphic involution satisfying $\sigma^*\Omega_2=\Omega_2$, where the action of $\sigma^*$ on the holomorphic two-form $\Omega_2$ on K3 has two $-1$ eigenvalues\footnote{Several explicit examples are given in \cite{K3Orientifold} for the case of K3 orbifolds.}. Compact internal spaces M$_4$ require, of course, the cancellation of the total RR charge as a consistency condition. The O-planes can carry RR charges and in many models it is necessary to add D-branes to ensure charge cancellation. For non-compact internal sector this constraint can in principle be relaxed.\\

Since the low energy limit of these models corresponds to supergravity theories on AdS$_3$, we could follow a different path: instead of directly dealing with orientifolds of strings living in these backgrounds, we could analyze the string compactification to $d=6$ before reducing to the AdS$_3$ vacuum solution, and then perform the corresponding Kaluza-Klein compactification on S$^3$. In the following, we briefly consider an orientifold of the torus compactification of type IIB superstring to $d=6$.

A torus compactification preserves all supersymmetry coming from the $d=10$ theory. Here one has two supercharges $Q_{\alpha}$, $\tilde{Q}_{\alpha}$, $\alpha=1,...,16$, one from the left-moving and one from the right-moving sector, which are both ten-dimensional spinors of the same chirality \cite{Friedan:1985ge}. These spinors decompose as
\begin{equation}
	Q_\alpha\rightarrow\ (Q^a_A, Q^{A\dot{a}}),\hspace{1cm} \tilde{Q}_\alpha\rightarrow\ (\tilde{Q}^a_A, \tilde{Q}^{A\dot{a}})
\end{equation}
where upper (lower) $A=1,..,4$ is an index for a chiral (antichiral) spinor of $\mathfrak{so}(1,5)$, while $a$ ($\dot{a}$) is a spinor index for the first (second) $\mathfrak{su}(2)$ contained in $\mathfrak{so}(4)\equiv \mathfrak{su}(2)\oplus \mathfrak{su}(2)$.
Hence, the low energy field theory is ${\cal N}=(2,2)$ $d=6$ supergravity.\\

Bosonic $d=10$ massless states of type IIB are:
\begin{itemize}
	\item	NS-NS sector: $G_{MN}, B_{MN}, \Phi$.
	\item	R-R sector: $C, C_{MN}, C_{MNPQ}$.
\end{itemize}
Under the split $M\rightarrow (m,i)$, where $m=0,...,5$, $i=6,...,9$, and neglecting dependence on internal coordinates, one gets the field content in $d=6$:
\begin{itemize}
	\item	25 scalars: one dilaton $\Phi$, sixteen moduli of $\mathbb{T}^4$ (ten from $G_{ij}$ and six from $B_{ij}$), one axion $C$, $b_2=6$ scalars from $C_{ij}$, and one scalar $C_{ijkl}$.
	\item	16 vectors: $G_{mi}$, $B_{mi}$, $C_{mi}$, $C_{mijk}$; each contributing 4 vectors.
	\item	5 antisymmetric tensors: one $B_{mn}$, one $C_{mn}$ and three from $C_{mnij}$.
	\item	Metric: $G_{mn}$
\end{itemize}

This is precisely the bosonic field content of ${\cal N}=(2,2)$ $d=6$ supergravity 
.\\

Now consider the orientifold of type IIB superstring on the torus $\mathbb{T}^4$ with orientifold group $\{1,\Omega_w\sigma\}$. The action of $\sigma$ on the spin fields (of the RNS formalism) allows one to deduce \cite{Berkooz:1996dw}
\begin{equation}\label{sigmaQ}
	\sigma Q^a_A = -Q^a_A, \hspace{1cm}\sigma Q^{A\dot{a}} = Q^{A\dot{a}}
\end{equation}
and similarly for the supercharges in the right-moving sector.
This means that only $Q^a_A -\tilde{Q}^a_A$ and $Q^{A\dot{a}}+\tilde{Q}^{A\dot{a}}$ are left invariant by $\Omega_w\sigma$; the low energy field theory is ${\cal N}=(1,1)$ $d=6$ supergravity coupled to matter multiplets.

To get the appropriate projection of massless states, notice that worldsheet supersymmetry requires that
\begin{equation}
	\sigma\psi^m = \psi^m, \hspace{1cm}\sigma\psi^i=-\psi^i\,,
\end{equation}
with the analogous action on right-moving $\tilde{\psi}^M$. Combining this with worldsheet parity reversal, one gets in the NS-NS sector:
\begin{equation}
	\Omega_w\sigma \Phi=\Phi,
	\hspace{1cm}\Omega_w\sigma G_{ij}=G_{ij}, 
	\hspace{1cm}\Omega_w\sigma B_{mi}=B_{mi}, 
	\hspace{1cm}\Omega_w\sigma G_{mn}=G_{mn}
\end{equation}
\begin{equation}
	\Omega_w\sigma B_{ij}=-B_{ij}, 
	\hspace{1cm}\Omega_w\sigma G_{mi}=-G_{mi},
	\hspace{1cm}\Omega_w\sigma B_{mn}=-B_{mn}
\end{equation}

The projection in the R-R sector requires one to know the action of $\sigma$ on the different components of the fermion vertex ${\mathcal{V}}_\alpha=e^{-\phi/2}S_{\alpha}$ (and $\tilde{\mathcal{V}}_\alpha=e^{-\tilde{\phi}/2}\tilde{S}_{\alpha}$). This was already given in (\ref{sigmaQ}); the left-moving supersymmetry charge is just the left-moving fermion vertex at zero momentum.
Notice also that ${\mathcal{V}}_{\alpha}$ and $\tilde{\mathcal{V}}_{\alpha}$ anticommute.

R-R states are obtained by tensoring ${\mathcal{V}}_\alpha\otimes\tilde{\mathcal{V}}_\beta$. The action of $\sigma$ is
\begin{equation}
	\sigma\ {\mathcal{V}}^a_A\tilde{\mathcal{V}}^b_B = {\mathcal{V}}^a_A\tilde{\mathcal{V}}^b_B,
	\hspace{1cm}\sigma\ {\mathcal{V}}^{A\dot{a}}\tilde{\mathcal{V}}^{B\dot{b}} = {\mathcal{V}}^{A\dot{a}}\tilde{\mathcal{V}}^{B\dot{b}},
\end{equation}
\begin{equation}
	\sigma\ {\mathcal{V}}^a_A \tilde{\mathcal{V}}^{B\dot{b}}= -{\mathcal{V}}^a_A \tilde{\mathcal{V}}^{B\dot{b}},
	\hspace{1cm}\sigma\ {\mathcal{V}}^{A\dot{a}}\tilde{\mathcal{V}}^b_B = -{\mathcal{V}}^{A\dot{a}}\tilde{\mathcal{V}}^b_B.
\end{equation}
From (linear combinations of) ${\mathcal{V}}^a_A\tilde{\mathcal{V}}^b_B$ and ${\mathcal{V}}^{A\dot{a}}\tilde{\mathcal{V}}^{B\dot{b}}$ one constructs the vertex operators of $C, C_{ij}$, $C_{ijkl}, C_{mn}$ and $C_{mnij}$, while ${\mathcal{V}}^a_A \tilde{\mathcal{V}}^{B\dot{b}}$ and ${\mathcal{V}}^{A\dot{a}}\tilde{\mathcal{V}}^b_B$ give vertex operators of $C_{mi}$ and $C_{mijk}$. Taking the action of $\Omega_w$, one gets
\begin{equation}
	\Omega_w\sigma C_{ij} =C_{ij}, \hspace{1cm}
	\Omega_w\sigma C_{mijk} =C_{mijk}, \hspace{1cm}
	\Omega_w\sigma C_{mn} =C_{mn},
\end{equation}
\begin{equation}
	\Omega_w\sigma C=-C, 
	\hspace{0.6cm}
	\Omega_w\sigma C_{ijkl} =-C_{ijkl}, \hspace{0.6cm}
	\Omega_w\sigma C_{mi} =-C_{mi},
	\hspace{0.6cm}
	\Omega_w\sigma C_{mnij} =-C_{mnij}.
\end{equation}

Therefore, the bosonic massless states of this orientifold of type IIB on $\mathbb{T}^4$ are
\begin{itemize}
	\item	17 scalars: one dilaton $\Phi$, ten scalars from $G_{ij}$ and six scalars from $C_{ij}$.
	\item	8 vectors: $B_{mi}$ and $C_{mijk}$; each contributing 4 vectors.
	\item	1 antisymmetric tensor: $C_{mn}$.
	\item	Metric: $G_{mn}$
\end{itemize}
This matches the field content of ${\cal N}=(1,1)$ $d=6$ supergravity coupled to 4 vector multiplets. The states projected out from the type IIB parent theory are thus 8 scalars $B_{ij}$, $C$ and $C_{ijkl}$, 8 vectors $G_{mi}$ and $C_{mi}$, and four antisymmetric tensors $B_{mn}$ and $C_{mnij}$. The spacetime fields corresponding to these states (together with their duals in $d=6$) must be absent in the supergravity analysis and, in particular, in all the equations of motion.

Notice that our orientifold projection does not kill $C_{mn}$, whose component $C_{\mu a}$ is crucial in order to get a consistent set of linearized equations as explained in section \ref{sec:spin1}.

If we now take the decompactification limit,
a continuum appears in the massless spectrum due to the presence of internal momenta (corresponding massive states become massless while winding states acquire infinite mass). 
The theory would in principle become effectively ten-dimensional, although the projection sketched before is still valid.
The question arises whether compactification induced by warping can lift this continuum and allow us to make contact with the $\mathbb{R}^4$ solutions presented in section \ref{sec:fibredsols}. Notice that the case with a single O5-plane seems more naturally related to the $R\rightarrow\infty$ limit of the torus solution; the O5-O5 solution seems more complicated from the string perspective.


\section{Conclusions}
\label{sec:conclusions}

The results of this work consist of two main parts, the construction of new AdS$_3$ solutions in type IIB supergravity that preserve small ${\cal N}=(4,0)$ supersymmetry, and work towards the computation of the spectra of their dual CFT$_2$s through the AdS-CFT correspondence.\\
~~\\
In the first part we consider a class of solutions, originally derived in \cite{Lozano:2019emq} which generalise the D1-D5 near horizon solution (AdS$_3\times$S$^3\times$CY$_2$ preserving ${\cal N}=(4,4)$) in two ways: i) First there is an additional CY$_2$ dependent warping consistent with co-dimension 4 sources ii) The 3-sphere becomes non trivially fibered over CY$_2$ in terms of a connection ${\cal A}$. These additions come at the cost of breaking supersymmetry to ${\cal N}=(4,0)$, but have the benefit of no longer requiring that CY$_2= \mathbb{T}^4$ or K3 for the internal space to be bounded.

We consider the case of CY$_2= \mathbb{R}^4$ in the most detail, constructing explicit solutions with non trivial ${\cal A}$ bounded between D5 and O5, O5 and O5 singularities as well as an intermediate case bounded between an O5 and a regular zero. In addition to this we show how to backreact D5 and O5 on AdS$_3\times$S$^3\times \mathbb{T}^4$, giving an explicit example with a single  O5 plane. These examples non trivially generalise the D1-D5 near horizon providing new examples where the internal metric is explicitly known.

Generalising out from this point we also establish how to construct solutions for generic  CY$_2$s that contain at least one U(1) isometry, with a view towards constructing solutions of more exotic Calabi-Yau manifolds in future. Finally we generalise the class of \cite{Lozano:2019emq} to provide an uplift of all solutions of minimal $d=5$ ungauged supergravity coupled to an Abelian vector multiplet. This provides new embeddings for (among other things) $d=5$ black-hole and black-string solutions into type IIB supergravity, opening a new window into microstate counting a la Strominger-Vafa \cite{Strominger:1996sh}.

For string Phenomenological reasons it is of course interesting to consider the D1-D5 near horizon on K3. This is an example where Calabi-Yau manifolds are known to exist but no closed form (or even realistically tractable series) expressions for their metrics is known. Despite this much progress has been made over the years considering string theory on K3 in various contexts. A glaring omission from our work is the backreaction of sources on AdS$_3\times$S$^3\times$K3. This should indeed be possible within the class of solution of \cite{Lozano:2019emq}, however the methods we employed to construct solutions required information about the metric on CY$_2$ to construct the connection ${\cal A}$. We have little doubt that with a more sophisticated approach the need to define a metric can be circumvented -- we leave this interesting problem for future work.\\
~\\
The main motivation behind the study of the above supergravity solutions is deriving new examples of the AdS$_3$/CFT$_2$ correspondence. For the reasons explained in the introduction, this topic remains a subject of active research. So, in the second part of the work, we made a few steps towards the identification of possible dual CFT$_2$. The main target of the present study was the superconformal multiplet containing the stress-energy tensor and the triplet R current. We identified the corresponding supergravity modes. In supergravity the R current mode appears mixed with a few other vector modes, for which we obtained the scaling dimensions of the dual CFT operators.

The presence of the stress-tensor superconformal multiplet was guaranteed by supersymmetry, so the most interesting part about our results is the specific form of the equation that describes the R current mode. In the present analysis this equation was derived for the simplest representative of the family of supergravity backgrounds, but it is possible to generalise it to the whole family and beyond, making it similar to the equation describing the stress-energy tensor mode, which in turn is valid for a very large class of compactifications. We expect that the R current mode equation is valid for a large class of backgrounds with extended supersymmetry. Further details about this matter will appear in the future publication~\cite{inprep}.

A somewhat unexpected behaviour was observed in the analysis of the non-BPS part of the spectrum, which is parallel for both the spin two and spin one sectors. We find that for some classes of backgrounds (D5-O5) the spectrum of non-BPS states is finite, bounded from above by a continuum of non-normalisable modes. This bound has a simple dependence on the compactification parameters, such as the number of D branes. In the simplest example of the background considered in section~\ref{sec:basicsol} there are no modes other than the BPS ones. We do not know the precise reasons for such a behaviour, but a possible explanation is the singularity of the background and breaking of the supergravity approximation close to the singular points.

The analysis of this paper is only a first step towards the identification of the dual theories. A desirable direction would be the full string theory analysis of the discussed compactifications. Only a qualitative discussion of the string approach was given in this work in section~\ref{sec:stringsAdS3}. We hope to come back with more specific results in the future.


\section*{Acknowledgments}
We are indebted to Thiago Fleury for extensive discussions and his collaboration on related projects. We also thank Alessandro Tomasiello and Kiril Hristov for useful correspondences. The work of ML was supported by CAPES, a Foundation of the Brazilian Ministry of Education and the Serrapilheira Institute (grant number Serra – R-2012-38185). She also thanks the Physics Department of Swansea University for the hospitality during the final stage of this project. NM is currently supported by grants from the Spanish government  MCIU-22-PID2021-123021NB-I00 and principality of Asturias  SV-PA-21-AYUD/2021/52177. He acknowledges support during earlier stages of this project from AEI-Spain (under project PID2020-114157GB-I00 and Unidad de Excelencia Mar\'\i a de Maetzu MDM-2016-0692), the Xunta de Galicia-Conseller\'\i a de Educaci\'on (Centro singular de investigaci\'on de Galicia accreditation 2019-2022, and project ED431C-2021/14), and by the European Union FEDER. The work of DM in this project was supported by the Russian Science Foundation grant {\#}21-12-00400. The work of LY was partially supported by the Simons Foundation award {\#}884966, the Association International Institute of Physics, and partially by Dr. Carlos Mafra's Royal Society University Research Fellowship.


\begin{appendix}

\section{The general form of CY$_2$ manifolds with a U(1) isometry}
\label{sec:gency2withu1appendix}

In this appendix we derive the local form of all CY$_2$ manifolds which contain a U(1) isometry. As we shall see there are two such classes, one governed by a Laplace equation in three dimensions, the other by a Toda equation.

Following \cite{Bah:2013qya} the most general complex metric in four dimensions containing a U(1) isometry, here $\partial_{\phi}$, can be written locally in the form
\beq
ds^2(\text{CY}_2)= e^{2C-2B}D\phi^2+ e^{2B}Dy^2+ e^{2\Delta+2B}(dx_1^2+dx_2^2)\,,\qquad  D\phi= d\phi+{\cal B}\,, \qquad Dy=dy+e^{-C}\tilde{\cal B}\,,
\eeq
where $({\cal B},\tilde{\cal B})$ have legs in $(x_1,x_2)$ only, but everything has functional support in $(y,x_1,x_2)$. The corresponding SU(2) structure takes the form
\beq
J\ =\  \frac{i}{2}\left(E_1\wedge \overline{E}_1+E_2\wedge \overline{E}_2\right)\,,\qquad \Omega\ =\  e^{-i n \psi} E_1\wedge E_2\,,
\eeq
where $n$ is constant and the complex vielbein $E_{1,2}$ are
\beq\label{eq:complexvielbein}
E_1\ =\  e^{C}D\psi+i e^{B-C}Dy\,,\qquad E_2 \ =\  e^{\Delta+B}(dx_1+i dx_2)\,.
\eeq
Imposing that $dJ=0$ one quickly extracts
\beq
\partial_{x_1}\tilde{\cal B}_{x_2}\ =\ \partial_{x_2}\tilde{\cal B}_{x_1}\,, \qquad \partial_y \tilde {\cal B}_{x_i}\ =\  \partial_{x_i}e^{C}\,.
\eeq
The first of these is an integrability condition, which we can solve locally in terms of a function $g=g(y,x_1,x_2)$ as
\beq
{\cal B}_{x_i}\ =\ \partial_{x_i}g\,,
\eeq
without loss of generality. The general solution to the second PDE is then 
\beq
e^{C}\ =\ \partial_y g \qquad \Rightarrow \qquad Dy\ =\  \frac{1}{\partial_y g}dg\,,
\eeq
which tells us we can actually take $g$ to be a coordinate, then use diffeomorphism invariance to fix
\beq
g\ =\ y\,,\qquad e^{C}\ =\ 1\,,\qquad Dy\ =\ dy.
\eeq
The remaining PDEs that follow from imposing $dJ=d\Omega=0$ are then simplified to
\begin{align}
&\partial_{x_i}{\Delta}=n \epsilon_{ij} {\cal B}_{x_j}\,,\qquad \partial_y\Delta=ne^{2B}\,,\qquad \partial_{y}{\cal B}_{x_i}=-\epsilon_{ij}\partial_{x_j}e^{2B},\label{eq:genericCY2U1PDEsa}\\[2mm]
&2e^{2\Delta}(n e^{4B}+\partial_y e^{2B})=-\epsilon_{ij}\partial_{x_i}{\cal B}_{x_j}\,.
\label{eq:genericCY2U1PDEsb}
\end{align}
This system contains exactly two physically distinct classes where CY$_2$ is defined in terms of a single PDE: For the first \eqref{eq:genericCY2U1PDEsa} can be solved without loss of generality as
\beq
n\ =\ 0\,,\qquad \Delta\ =\ 0\,,\qquad e^{2B}\ =\ \partial_y H\,,\qquad {\cal B}\ =\ -\epsilon_{ij}\partial_{x_j} H\,.
\eeq
then \eqref{eq:genericCY2U1PDEsb} is reduced to a Laplacian on $\mathbb{R}^3$
\beq
(\partial^2_{x_1}+\partial_{x_2}^2+\partial_{y}^2) H\ =\ 0\,.
\eeq
For the second, \eqref{eq:genericCY2U1PDEsa} is solved without loss of generality as
\beq
n\ =\ 1\,,\qquad {\cal B}_{x_i}\ =\ -\epsilon_{ij}\partial_{x_j}\Delta\,,\qquad  e^{2B}\ =\ \partial_y\Delta
\eeq
and \eqref{eq:genericCY2U1PDEsb} reduces to a Toda equation
\beq
2(\partial_{x_1}^2+\partial_{x_2}^2)\Delta+ \partial_{y}e^{2\Delta}\ =\ 0\,.
\eeq 
These leads to the two classes of CY$_2$ manifolds we quote in section \ref{sec:circlefibration}.

\section{The ${\cal N}=4$ superalgebra}

There are two kinds of $\mathcal{N}=4$ superconformal algebras, the small and large ones. Both types appeared in this work and we review them in this Appendix. 

\subsection{The small ${\cal N}=4$ superalgebras}
\label{sec:smallN4}

The $\mathcal{N}=4$ \emph{small} superconformal algebra has two sets of bosonic generators, $L_m$ and $T_m^i$, and two sets of fermionic ones, $G_r^a$ and $\bar{G}_r^a$. 
The indices assume values $a,b= 1,2$, $i= 1,2,3$, and $r,s$ are integers (half-integers) in the Ramond (Neveu-Schwarz) sector. The algebra is 
\begin{align}
&[L_m,L_n]=(m-n)L_{m+n}+\frac{k}{2}m(m^2-1)\delta_{m+n,0},\\
&[T_m^i,T_n^j]=i\epsilon^{ijk}T^k_{m+n}+\frac{1}{2}km\delta_{m+n,0}\delta^{ij},\\
&\{G_r^a,G_s^b\}=\{\bar{G}_r^a,\bar{G}_s^b\}=0,\\
&\{G_r^a,\bar{G}_s^b\}=2\delta^{ab}L_{r+s}-2(r-s)\sigma_{ab}^iT^i_{r+s}+
\frac{1}{2}k(4r^2-1)\delta_{r+s,0}\delta^{ab},\label{GG}\\
&[L_m,T_n^i]=-nT^i_{n+m},\qquad\qquad\! [L_m,G_r^a]=\left(\frac{m}{2}-r\right)G_{m+r}^a,\qquad
[L_m,\bar{G}_s^a]=\left(\frac{m}{2}-s\right)\bar{G}_{m+s}^a,\\
&[T_m^i,G_r^a]=-\frac{1}{2}\sigma_{ab}^iG_{m+r}^b,\qquad
[T_m^i,\bar{G}_s^a]=\frac{1}{2}\sigma_{ab}^{i\,*}\bar{G}_{m+s}^b,
\end{align}
where $\epsilon^{ijk}$ is the totally antisymmetric tensor with $\epsilon^{123}=1$ and $\sigma^i_{ab}$ are the Pauli matrices.
Notice that in a unitary representation $k$ must be a non-zero positive integer \cite{Eguchi:1987sm}.
One can find the algebra written in the OPE language in \cite{Matsuda:1988qf} so that
$L_m$, $T_m^i$, $G_r^a$, and $\bar{G}_s^a$, are the Fourier components of the energy-momentum tensor, the SU(2) current operators, and the supercharges, respectively.

\subsubsection{Long and short multiplets in the NS sector}
\label{NS}

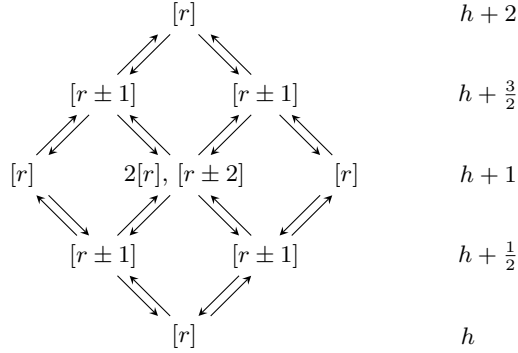
\begin{figure}[h]
\begin{center}
\begin{tikzpicture}[scale=0.8]
     \draw[-stealth](0.1,0) -- (-2/3+0.1,2/3); 
    \draw[stealth-](-0.1,0) -- (-2/3-0.1,2/3); 
    \draw[stealth-](-4/3+0.1,4/3) -- (-2+0.1,2); 
    \draw[-stealth](-4/3-0.1,4/3) -- (-2-0.1,2); 
    \draw[stealth-](-2+0.1,2+2/3) -- (-2+2/3+0.1,2+4/3); 
    \draw[-stealth](-2-0.1,2+2/3) -- (-2+2/3-0.1,2+4/3); 
    \draw[stealth-](-2+4/3+0.1,4) -- (0.1,4+2/3); 
    \draw[-stealth](-2+4/3-0.1,4) -- (-0.1,4+2/3); 
    \draw[stealth-](2/3+0.1,4+2/3) -- (4/3+0.1,4);
    \draw[-stealth](2/3-0.1,4+2/3) -- (4/3-0.1,4);
    \draw[stealth-] (2+0.1,4-2/3) -- (8/3+0.1,4-4/3); 
    \draw[-stealth] (2-0.1,4-2/3) -- (8/3-0.1,4-4/3); 
    \draw[stealth-] (4/3+4/3+0.1,4-4/3-2/3) -- (4/3+2/3+0.1,4-4/3-4/3); 
    \draw[-stealth] (4/3+4/3-0.1,4-4/3-2/3) -- (4/3+2/3-0.1,4-4/3-4/3); 
    \draw[stealth-] (4/3+0.1,2-4/3) -- (2/3+0.1,0); 
    \draw[-stealth] (4/3-0.1,2-4/3) -- (2/3-0.1,0); 
    \draw[-stealth] (4/3+0.1,4-4/3-4/3) -- (2/3+0.1,4-4/3-2/3); 
    \draw[stealth-] (4/3-0.1,4-4/3-4/3) -- (2/3-0.1,4-4/3-2/3); 
    \draw[stealth-] (4/3-2+0.1,4-4/3-4/3) -- (0.1,4-4/3-2/3);
    \draw[-stealth] (4/3-2-0.1,4-4/3-4/3) -- (-0.1,4-4/3-2/3);
    \draw[-stealth] (0.1,4-4/3) -- (-2/3+0.1,4-2/3);
    \draw[stealth-] (-0.1,4-4/3) -- (-2/3-0.1,4-2/3);
    \draw[-stealth] (-2/3+2+0.1,4-2/3) -- (-4/3+2+0.1,4-4/3);
    \draw[stealth-] (-2/3+2-0.1,4-2/3) -- (-4/3+2-0.1,4-4/3);
    \node[] at (-2-1/3,2+1/3) {\small $[r]$};
    \node[] at (1/3,4-4/3-1/3) {\small $2 [r],\, [r\pm 2]$};
    \node[] at (1/3+5,4-4/3-1/3) {\small $h+1$};
    \node[] at (1/3,-1/3) {\small $[r]$};
    \node[] at (5,-1/3) {\small $h$};
    \node[] at (-1,1) {\small $[r\pm 1]$};
    \node[] at (1/3,4-4/3-1/3+8/3) {\small $[r]$};
    \node[] at (1/3+5,4-4/3-1/3+8/3) {\small $h+2$};
    \node[] at (-1,1+8/3) {\small $[r\pm 1]$};
    \node[] at (1/3+5,1+8/3) {\small $h+\frac{3}{2}$};
    \node[] at (-1+8/3,1+8/3) {\small $[r\pm 1]$};
    \node[] at (1/3+8/3,4-4/3-1/3) {\small $[r]$};
    \node[] at (-1+8/3,1) {\small $[r\pm 1]$};
    \node[] at (1/3+15/3,1) {\small $h+\frac{1}{2}$};
\end{tikzpicture}
\end{center}
\caption{Schematic representation of the long-multiplet $\mathcal{L}_r$, see   \cite{Kos:2018glc} for more details. The arrows going to the left (right) represent the action of $G^a_{-\frac{1}{2}}$ $\left(\bar{G}^b_{-\frac{1}{2}}\right)$. In the figure only the $SU(2)$ indices are shown. Operators in the same line have the same conformal dimension $h$ and the dimension grows by half when moving to the next line above.  }

\label{fig:smultiplet}
\end{figure}

The superconformal primary operators are defined as those which are annihilated by all positive Fourier modes and are eigenstates of the zero modes, that is,
\begin{align}
&L_{n>0}\ket{\mathcal{O}_{h,r}}=G^a_{r>0}\ket{\mathcal{O}_{h,r}}=
\bar{G}^a_{s>0}\ket{\mathcal{O}_{h,r}}=T^i_{m>0}\ket{\mathcal{O}_{h,r}}=0,\nonumber\\
&L_0\ket{\mathcal{O}_{h,r}}=h\ket{\mathcal{O}_{h,r}},\qquad
T_0^i\ket{\mathcal{O}_{h,r}}= t^i \ket{\mathcal{O}_{h,r}}.
\end{align}
The global long-multiples are obtained by acting repeatedly with $G_{-1/2}^a$ and $\bar{G}_{-1/2}^a$ in $| \mathcal{O}_{h,r} \rangle$ until they annihilate it. 
Schematically, the long multiplets $\mathcal{L}_r$ contain the set of operators of figure~\ref{fig:smultiplet}. 

It is possible to have all the multiplet components obeying the conformal primary condition $L_1 \ket{\mathcal{O}_{h^{\prime},r^{\prime}}}=0$, by shifting some of them by a correction term,  suitably changing the basis. All the corrections for any component of $\mathcal{L}_0$ and $\mathcal{L}_1$ can be found in \cite{Kos:2018glc}. Altogether, a long multiplet $\mathcal{L}_{r>1}$ consists of a set of sixteen\footnote{$\mathcal{L}_0$ has ten and $\mathcal{L}_1$ has fifteen operators.} (quasi)conformal primary operators built from $|\mathcal{O}_{h,r}\rangle.$

Imposing unitarity we obtain
\begin{align}\label{short}
\bra{\mathcal{O}_{h,r}}\{ G^a_{\frac{1}{2}},\bar{G}^b_{-\frac{1}{2}}\}
\ket{\mathcal{O}_{h,r}}=\bra{\mathcal{O}_{h,r}}2\delta^{ab}L_0-2\sigma_{ab}^iT_0^i\ket{\mathcal{O}_{h,r}}=
2\left(h-\frac{r}{2}\right)\geq 0.
\end{align}
The equality $h=r/2$ saturating the lower bound in \eqref{short} is called the shortening condition (SC). One calls a short multiplet the set of states consisting of the bottom component, which is a superconformal primary $|\mathcal{O}_{h,r}\rangle$ obeying SC, and its decedents. We will denote the short multiplets by $\mathcal{A}_r$. The shortening condition for $h=r=0$ is obeyed if $h=0$. The vacuum is the only state with $h=0$, and the correspondent operator is the identity. Furthermore, in this case the identity is the only superconformal primary operator, then 
$\mathcal{A}_0= \{ \textbf{1}\}$. 

In order to construct $\mathcal{A}_1$ we have to start from a state which has $h=1/2$ to satisfy the SC. One can see that we can only construct two other operators with lower dimension $SU(2)$ representations, therefore
\begin{equation}
\mathcal{A}_1= \{
|\mathcal{O}_{\frac{1}{2},\frac{1}{2}} \rangle, G_{-\frac{1}{2}} |\mathcal{O}_{\frac{1}{2},\frac{1}{2}} \rangle, 
\bar{G}_{-\frac{1}{2}} |\mathcal{O}_{\frac{1}{2},\frac{1}{2}} \rangle\} \, . 
\end{equation}
For $r>1$ we obtain 
\begin{equation}
\mathcal{A}_r=
\left\{|\mathcal{O}_{\frac{r}{2},r}\rangle,
|\mathcal{O}_{\frac{r+1}{2},r-1}\rangle,
|\mathcal{O}_{\frac{r+1}{2},r-1}\rangle,
|\mathcal{O}_{\frac{r+1}{2},r-2} \rangle  \right\},
\end{equation}
where $|\mathcal{O}_{r/2,r} \rangle$ is a superconformal primary operator. It is easy to see that each small internal diamond in figure~\ref{fig:smultiplet} forms a short multiplet when we impose the SC for their bottom component. The long multiplets $\mathcal{L}_r$ can be decomposed in a direct sum of short multiplets.

Any superconformal theory has a short multiplet containing the stress-energy tensor, which is nothing but a generator of the superconformal algebra. The latter should commute with $G^a_{-\frac{1}{2}}$ and $\bar{G}^a_{-\frac{1}{2}}$, so we expect that it to be the top component of the multiplet. In particular, any short multiplet with bottom component $[r=2]$ has a top component with $(h,r)=(2,0)$. In the case of the stress-energy tensor the lowest component of the short superconformal multiplet is the SU(2) R current operator with $(h,r)=(1,2)$. The gravity mode corresponding to the latter operator was identified in section~\ref{sec:spin1}.

\subsubsection{The R sector}
As it was said before, in the R sector the indices $r$ and $s$ of the fermionic operators assume integer values. In this section we will denote the highest-weight states by $\ket{\mathcal{O}_{h,r}'}$, with $r$ being the $T_0^3$ eigenvalue \cite{Eguchi:1987wf}:
\begin{align}\label{hws}
&L_{n>0}\ket{\mathcal{O}_{h,r}'}=G^a_{m>0}\ket{\mathcal{O}_{h,r}'}=
\bar{G}^a_{m>0}\ket{\mathcal{O}_{h,r}'}=T^i_{a>0}\ket{\mathcal{O}_{h,r}'}=0,\\
&G_0^2\ket{\mathcal{O}_{h,r}'}=\bar{G}_0^1\ket{\mathcal{O}_{h,r}'}=T_0^+\ket{\mathcal{O}_{h,r}'}=0, \quad 
L_0\ket{\mathcal{O}_{h,r}'}=h\ket{\mathcal{O}_{h,r}'},\quad T_0^3\ket{\mathcal{O}_{h,r}'}=r\ket{\mathcal{O}_{h,r}'},
\end{align}
where $T_m^\pm=T_m^1+iT_m^2$. One can see that the operators $G_0^2$ and $\bar{G}_0^1$ increase $r$. This time, 
the shortening condition is derived from the zero $G$ modes 
\begin{equation}\label{nsc}
\bra{\mathcal{O}_{h,r}'}\{G_0^a,\bar{G}_0^b\}\ket{\mathcal{O}_{h,r}'}=\delta^{ab}\left(2h-\frac{k}{2}\right)\geq 0.
\end{equation}

We call massless and massive representations the two cases
\begin{align}
\begin{cases}
&\text{massless}:\qquad h=\frac{k}{4},\qquad r=0,\frac{1}{2},1,...,\frac{k}{2},\\
&\text{massive}:\,\qquad h>\frac{k}{4},\qquad r=\frac{1}{2},1,\frac{3}{2},...,\frac{k}{2}.
\end{cases}
\end{align}
We saw that in the NS sector long multiplets can be written in terms of short multiplets, similarly, massive irreducible representation can be express in terms of massless irreducible ones \cite{Eguchi:1987sm}.

Considering the algebra of the zero modes, one can verify that $\bar{G}_0^1$ and $G_0^2$ increase $r$ by $1/2$, while $G_0^1$ and $\bar{G}_0^2$ decrease $r$ by the same value. Also, $T_0^+$ increases $r$ by 1 and $T_0^-$ decreases $r$ by 1. We define $A_1$ as the state with the maximum $T_0^3$ eigenvalue in the representation, $T_0^3A_1=RA_1$, then $G_0^2A_1=\bar{G}_0^1A_1=T_0^+A_1=0$. 
In addition, $L_0 A_1 = H A_1$. We define $B_1=G_0^1A_1$, $\bar{B}_1=\bar{G}_0^2A_1$, and the operators
\begin{align}\label{w0w1}
w_0=(T_{-1}^+)^{k-2l+1},\qquad w_1=(T_0^+)^{2l-1}\left[\bar{G}_0^2G_0^1-\frac{1}{l}\left(H-\frac{k}{4}\right)T_0^-\right].
\end{align}
Then, starting from $A_1$ we construct the full global supermultiplet in the massive\footnote{In the massless representation we exchange $w_1$ by $w_1=(T_0^+)^{2l-1}$.} representation by applying alternately $w_0$ and $w_1$ in $A_1$ (see $\alpha$- and $\beta$-series in \cite{Eguchi:1987wf}). The resulting supermultiplet is composed by the following states

\begin{enumerate}
    \item $A_1$ $A_2,\, A_3,...,A_{2R+1}$.
\item $B_1, \ldots, B_{2R}$.
\item $\bar{B}_1, 
\ldots, \bar{B}_{2R}$.
\item
$C_1=\bar{G}_0^2G_0^1A_1-
\frac{1}{2R}\left(2H-\frac{k}{2}\right)T_0^-A_1, \ldots, C_{2R-1}$. \end{enumerate}

It is important to note that in the massless representation the highest-weight states are annihilated not only by $\bar{G}_0^1$ and $G_0^2$, but also by $G_0^1$ and $\bar{G}_0^2$, what means that in such representation either all states are bosonic or all of them are fermionic depending on the grade of $A_1$.

\subsection{The large $\mathcal{N}=4$ superconformal algebra}
\label{sec:largeN4}

In the previous section we constructed the supermultiplets of the small ${\mathcal N}=4$ superconformal algebra. For some of the backgrounds discussed in this work, such as AdS$_3\times $S$^3\times $S$^3\times\mathbb{R}$ preserve the \emph{large} ${\mathcal N}=4$ superconformal algebra, which is defined by
\begin{align}
&[L_m,L_n]=(m-n)L_{m+n}+\frac{c}{12}
(m^3-m)\delta_{m+n,0},\\
&[U_m,U_n]=\frac{k^++k^-}{2}m\delta_{m,-n},\\
&[A_m^{\pm,i},A_n^{\pm,j}]=\frac{k^\pm}{2}m\delta^{ij}\delta_{m,-n}+i\epsilon^{ijl}A_{m+n}^{\pm,l},\\
&\{Q_r^a,Q_s^b\}=\frac{k^++k^-}{2}\delta^{ab}\delta_{r,-s},\\
&\{G_r^a,G_s^b\}=\frac{c}{3}\left(r^2-\frac{1}{4}\right)\delta^{ab}\delta_{r,-s}+2\delta^{ab}L_{r+s}+
4(r-s)\left[\gamma i\alpha_{ab}^{+i}A_{r+s}^{+,i}+(1-\gamma)i\alpha_{ab}^{-i}A_{r+s}^{-,i}\right],\\
&[L_m,V_n]=[(h(V)-1)m-n]V_{m+n},\label{Vn}\\
&[U_m,G_r^a]=mQ^a_{m+r}\\
&[U_m,A_n^{\pm,i}]=[U_m,Q_r^a]=0,\\
&[A_m^{\pm,i},Q_r^a]=i\alpha^{\pm i}_{ab}Q^b_{m+r},\\
&[A_m^{\pm,i},G_r^a]=i\alpha_{ab}^{\pm i}G^b_{m+r}\mp\frac{2k^\pm}{k^++k^-}m\alpha_{ab}^{\pm i}Q^b_{m+r},\\
&\{Q^a_r,G^b_s\}=2\alpha_{ab}^{+i}A^{+,i}_{r+s}-2\alpha_{ab}^{-i}A^{-,i}_{r+s}+\delta^{ab}U_{r+s},
\end{align}
where
\begin{equation}
\gamma=\frac{k^-}{k},\qquad c=\frac{6k^+k^-}{k},\qquad k=k^++k^-,
\end{equation}
and in \eqref{Vn} $V_n$ is either $U$, $A^{\pm,i}$, $Q^a$, or $G^a$, and $h$ is its respective conformal weight ($h[U]=1$, $h[A^{\pm,i}]=1$, $h[Q^a]=\frac{1}{2}$, and $h[G^a]=\frac{3}{2}$ \cite{Gaberdiel:2013vva}). In a unitary representation, both $k^+$ and $k^-$ are positive integers \cite{Gunaydin:1988re}. The matrices $\alpha_{ab}^{\pm i}$ are the $\mathfrak{so}(4)$ matrix representations of the generators
\begin{align}
\alpha_{ab}^{\pm i}=\frac{1}{2}\left(\pm\delta_{ia}\delta_{b0}\mp\delta_{ib}\delta_{a0}+\epsilon_{iab}
\right),
\end{align}
and they obey
\begin{align}
[\alpha^{\pm i},\alpha^{\pm j}]=-\epsilon^{ijl}\alpha^{\pm l},\qquad
[\alpha^{+ i},\alpha^{-j}]=0,\qquad \{\alpha^{\pm i},\alpha^{\pm j}\}=-\frac{1}{2}\delta^{ij}.
\end{align}
 The indices $a$ and $b$ run from 0 to 3, $i,j,l$ run from 1 to 3, $m,n\in \mathbb{Z}$,  $r,s\in\mathbb{Z}$ in the R sector and $r,s\in\frac{1}{2}\mathbb{Z}$ in the NS sector. Together the generators $L_0,\,L_{\pm1},\,G^a_{\pm\frac{1}{2}}$, and $A_0^{\pm,i}$ form the global superalgebra $D(2,1|\alpha)$ \cite{Gaberdiel:2013vva}.

We will denote the highest-weight states by $\ket{h,l^+,l^-,u}$, where $h$ and $u$ are the $L_0$ and $U_0$ eigenvalues, respectively, while $l^\pm$ are the su(2) spins\footnote{The fields $A^{\pm,i}$ and $U$ generate the current algebras
\begin{equation}
\mathfrak{su}(2)_{k^+}\oplus\mathfrak{su}(2)_{k^-}\oplus\mathfrak{u}(1).
\end{equation}} \cite{Gaberdiel:2013vva, Gunaydin:1988re}:
\begin{align}
&-(A^{+i}A^+{}_i)_0\ket{h,l^+,l^-,u}=l^+(l^++1)\ket{h,l^+,l^-,u},\\
&-(A^{-i}A^-{}_i)_0\ket{h,l^+,l^-,u}=l^-(l^-+1)\ket{h,l^+,l^-,u}.
\end{align}
They are annihilated by all the positive modes and by $A_0^{\pm+}$. The spins $l^\pm$ can assume the values
\begin{equation}
l^\pm=0,\frac{1}{2},1,...,\frac{1}{2}(k^\pm-1),
\end{equation}
that is, the unitarity bound is given by
\begin{equation}
l^\pm\leq\frac{1}{2}(k^\pm-1).
\end{equation}
Moreover, it is possible to show that the BPS bound is given by 
\begin{equation}
h \geq \frac{1}{k_+ + k_-}    
[k^+ j^- +
k^- j^+ + u^2 +(j^+ - j^-)^2] \, . 
\end{equation}

Once that a representation in the NS sector is constructed, one automatically has a representation in the R sector thanks to the equivalence \cite{Gunaydin:1988re}
\begin{align}
(h,l^+,l^-,u)_{NS}\equiv \left(h-l^++\frac{k^-}{4},l^+,\frac{k^-}{2}-l^-,u\right)_R.
\end{align}


\section{Scalar spherical harmonics on S$^3$}
\label{sec:harmonics}

As it shall be most useful to us, we shall give the scalar spherical harmonics on the 3-sphere in  Hopf fibration coordinates (see also~\cite{Lachieze-Rey:2005nyz}). In these coordinates the metric on the sphere is
\beq
ds^2(\text{S}^3) \ = \ \frac{1}{4}\bigg(d\theta^2+\sin^2\!\theta d\phi^2+ (d\psi+\cos\theta d\phi)^2\bigg),
\eeq
where $0\leq\theta<\pi$, $0\leq\phi<2\pi$, $0\leq\psi<4\pi$. 

Since the isometry group of S$^3$ is ${\rm SO(4)}\simeq{\rm SU(2)}\times{\rm SU(2)}$, the spherical harmonics can be labelled by a pair of spins $\ell_1$ and $\ell_2$ and their projections $m_1$ and $m_2$. Scalar harmonics correspond to the case $\ell_1=\ell_2=\ell$, while harmonics of spin $s$ have $s=|\ell_1-\ell_2|$.

The complex form of the scalar harmonics is given by
\beq
T_{\ell;m_1,m_2}\ = \ C_{\ell;m_1,m_2}e^{i (m_1\psi+m_2\phi)}\cos^{m_1+m_2}\!\left(\frac{\theta}{2}\right)\sin^{m_1-m_2}\!\left(\frac{\theta}{2}\right)P^{(m_1-m_2,m_1+m_2)}_{\frac{\ell}{2}-m_1}(\cos\theta)\,,
\eeq
where $P^{(\alpha,\beta)}_n(x)$ are Jacobi polynomials which solve the ODE
\beq
(1-x^2)P''+(\beta-\alpha-(\alpha+\beta+2)x)P'+n(n+\alpha+\beta+1)P \ = \ 0.
\eeq
Regularity requires that
\beq
\ell \ =\ 0,1,2,3,\ldots,\qquad  -\frac{\ell}{2}\leq m_1,m_2\leq\frac{\ell}{2}\,,
\eeq
where $m_1,m_2$ increase in integer increments making them  integer when $\ell$ is even and half integer when $\ell$ is odd. Thus we have
\begin{multline}
\int_0^{\pi}d\theta\ \sin\theta \cos^{2(m_1+m_2)}\!\left(\frac{\theta}{2}\right)\sin^{2(m_1-m_2)}\!\left(\frac{\theta}{2}\right)P^{(m_1-m_2,m_1+m_2)}_{\frac{\ell}{2}-m_1}(\cos\theta)P^{(m_1-m_2,m_1+m_2)}_{\frac{\ell'}{2}-m_1}(\cos\theta) \\ 
=\ \frac{2}{(\ell+1)}\frac{(\frac{\ell}{2}+m_2)!(\frac{\ell}{2}-m_2)!}{(\frac{\ell}{2}+m_1)!(\frac{\ell}{2}-m_1)!}\delta_{\ell\ell'}\,,
\end{multline}
The coefficients are then fixed as
\beq
C_{l;m_1,m_2}\ =\ \sqrt{\frac{\ell+1}{2\pi^2}}\sqrt{\frac{(\frac{\ell}{2}+m_1)!(\frac{\ell}{2}-m_1)!}{(\frac{\ell}{2}+m_2)!(\frac{\ell}{2}-m_2)!}}\,,
\eeq
so that
\beq
\int_{\text{S}^3} T_{\ell;m_1,m_2}T_{\ell';m_1',m_2'}\ d\text{vol}(\text{S}^3) \ = \ \delta_{\ell \ell'}\delta_{m_1 m_1'}\delta_{m_2 m_2'}\,,
\eeq
where $d\text{vol}(\text{S}^3)=\frac{1}{8}\sin\theta d\theta d\phi d\psi$. A solution to Laplace's equation on $\mathbb{R}^4=(\rho,\text{S}^3)$ is then given by
\beq
f\ =\ \sum_{\ell=0}^{\infty}\sum_{m_1,m_2=-\frac{\ell}{2}}^{\frac{\ell}{2}} c_{\ell;m_1,m_2} \rho^\ell\, T_{\ell;m_1,m_2}(\theta,\phi,\psi).
\eeq
Defining the SU(2) left and right invariant 1-forms
\be
\begin{array}{ll}
L_1+i L_2 \ =\ e^{i\psi}(i d\theta+\sin\theta d\phi)\,, & L_3 \ = \ d\psi+\cos\theta d\phi\,, \\
& \\
R_1+i R_2\ =\ e^{-i\phi}(i d\theta-\sin\theta d\psi)\,, & R_3\ = \ d\phi+\cos\theta d\psi 
\end{array}
\ee
and their dual Killing vectors defined as $(K^{L/R})_i^{\mu}=\frac{1}{4} g^{\mu\nu}(R_i/L_i)_{\mu}$, 
i.e.
\be
\begin{array}{ll}
K^{L}_1+i K^{L}_2 \ =\  e^{-i \phi}(i \partial_{\theta}+ \cot\theta \partial_{\phi}- \csc\theta\partial_{\psi})\,, & K^L_3 \ =\ \partial_{\phi}\,, \\
& \\
K^{R}_1+i K^{R}_2 \ =\  e^{i \psi}(i \partial_{\theta}- \cot\theta \partial_{\psi}+ \csc\theta\partial_{\phi}) \,, & K^R_3\ =\ \partial_{\psi}\,. 
\end{array}
\ee
One can show that 
\be
\begin{array}{ll}
\nabla^2_{\text{S}^3}T_{\ell;m_1,m_2} \ = \ -\ell(\ell+2)T_{\ell;m_1,m_2}\,, & \\
& \\
{\cal L}_{K^L_1\mp i K^L_2}T_{\ell;m_1,m_2}  = -i\sqrt{\frac{\ell}{4}(l+2)- m_2(m_2\pm 1)}T_{\ell;m_1,m_2\pm 1}, & {\cal L}_{K^L_3}T_{\ell;m_1,m_2} = i m_2T_{\ell;m_1,m_2},\\
& \\
{\cal L}_{K^R_1\pm i K^R_2}T_{\ell;m_1,m_2} = -i \sqrt{\frac{\ell}{4}(l+2)- m_1(m_1\pm 1)}T_{\ell;m_1\pm 1,m_2}, & {\cal L}_{K^R_3}T_{\ell;m_1,m_2} = i m_1T_{\ell;m_1,m_2}.
\end{array}
\ee
Note also that
\beq
4{\cal L}_{K^L_i}{\cal L}_{K^L_i}\ = \ 4{\cal L}_{K^R_i}{\cal L}_{K^R_i}\ = \ \nabla^2_{\text{S}^3}\,.
\eeq
When we parameterise $\mathbb{R}^4$ in terms of $(\rho,\text{S}^3)$ as we do in the generalised solutions, then a vector like
\beq
{\cal A}\ =\  \frac{\rho^2}{4}\sum_i c_i L_i
\eeq
is such that
\beq
|{\cal A}|^2\ =\  \frac{\rho^2}{4}(c_1^2+c_2^2+c_3^2)\,,\qquad \nabla_{{\cal A}}\ =\ {\cal L}_{\cal A}\ = \ \sum_i c_i {\cal L}_{K^R_i}\,,
\eeq
which are the quantities appearing in \eqref{FibSpin2}.

\end{appendix}

\end{document}